\documentclass[aps,eqsecnum,preprint,floats,epsf,epsfig,nofootinbib]{revtex4}
\usepackage{graphicx}
\usepackage{epsfig,graphicx,color,appendix}

\begin{document}
\def\be{\begin{eqnarray}}
\def\en{\end{eqnarray}}
\def\non{\nonumber}
\def\la{\langle}
\def\ra{\rangle}
\def\pp{{\prime\prime}}
\def\nc{N_c^{\rm eff}}
\def\vp{\varepsilon}
\def\hep{\hat{\varepsilon}}
\def\a{{\cal A}}
\def\B{{\cal B}}
\def\c{{\cal C}}
\def\d{{\cal D}}
\def\e{{\cal E}}
\def\p{{\cal P}}
\def\t{{\cal T}}
\def\up{\uparrow}
\def\dw{\downarrow}
\def\vma{{_{V-A}}}
\def\vpa{{_{V+A}}}
\def\smp{{_{S-P}}}
\def\spp{{_{S+P}}}
\def\J{{J/\psi}}
\def\ov{\overline}
\def\Lqcd{{\Lambda_{\rm QCD}}}
\def\pr{{Phys. Rev.}~}
\def\prl{{ Phys. Rev. Lett.}~}
\def\pl{{ Phys. Lett.}~}
\def\np{{ Nucl. Phys.}~}
\def\zp{{ Z. Phys.}~}
\def\lsim{ {\ \lower-1.2pt\vbox{\hbox{\rlap{$<$}\lower5pt\vbox{\hbox{$\sim$}
}}}\ } }
\def\gsim{ {\ \lower-1.2pt\vbox{\hbox{\rlap{$>$}\lower5pt\vbox{\hbox{$\sim$}
}}}\ } }

\font\el=cmbx10 scaled \magstep2{\obeylines\hfill March, 2011}

\vskip 1.5 cm

\centerline{\large\bf Decay constants and
 form factors of $s$-wave
and $p$-wave mesons}
 \centerline{\large\bf in the covariant light-front quark model}
\bigskip
\centerline{\bf R. C. Verma}
\medskip
\centerline{Institute of Physics, Academia Sinica, Taipei, Taiwan, 11529} 
\centerline{Permanent Address: Department of Physics, Punjabi University, Patiala, India, 147002}
\email{rcverma@gmail.com}
\medskip
\medskip
\medskip
\medskip
\medskip

\bigskip
\bigskip
\centerline{\bf Abstract}
\bigskip
\small

We reanalyze the decay constants of $s$-wave and $p$-wave mesons and
$D, B \rightarrow M$ form factors, where $M$ represents a pseudoscalar meson,
 a vector meson, a scalar meson,
 or an axial vector meson within a covariant
light-front quark model. The parameter $\beta$ for wave-functions of most of $s$-wave
 mesons and of a few axial-vector mesons are fixed with latest experimental
 information, wherever available or using the
  lattice calculations. The treatment of masses and mixing angles for
  strange axial vector mesons is improved for the purpose. We extend our analysis
    to determine the form factors appearing in the transition of
    $D_s, B_s\rightarrow M$ transitions, and to the isoscalar
    final state mesons. Numerical results of the form factors for
transitions between a heavy pseudoscalar meson and an $s$-wave or
$p$-wave light meson and their momentum dependence are presented in
detail. Further, their sensitivity
 to uncertainties of $\beta$ parameters of the initial as well
  as the final mesons is investigated. Some experimental measurements of the
  charmed and bottom meson decays  are employed to compare the decay constants
  and transition form factors obtained in this and other works.


\eject

\section{Introduction}

In the previous work \cite{Cheng04}, various $P \rightarrow M$ form factors,
where $P$ represents a heavy pseudoscalar meson $(D$ or $B)$, and $M$
represents either $s$-wave or low-lying $p$-wave meson, were calculated
within the framework of the covariant light-front (CLF)
 approach. This formalism preserves the Lorentz covariance in the
  light-front framework and has been applied successfully to describe
  various properties of pseudoscalar and vector mesons \cite{Jaus90,Jaus91,Jaus99}. The
  analysis of the covariant light-front quark model to transitions of the
   charmed and bottom mesons was extended to even parity, $p$-wave mesons \cite{Cheng04}.
   Recently, the CLF approach has also been used to the studies of the
quarkonia \cite{Shen08, Hwang07}, the $p$-wave meson emitting decays of the bottom mesons
 \cite{Cheng10} and the $B_c$ system \cite{Wang09} and so on.
    In the present work, we update our results for $D$ and $B$ meson form factors, and
   extend this analysis to determine the form factors appearing in the
       $D_s , B_s\rightarrow M$ transitions, and to the flavor-diagonal
       final state mesons $M$. Experimental measurements of the decays of the $\tau$ lepton,
       pseudoscalar and vector mesons are employed to determine the decay
   constants, which in turn fix the shape parameters, $\beta $, of the respective mesons.
   For a few cases, the decay constants estimated by lattice calculations
    have been used for this purpose. We have now used the improved estimation
    of the $K_{1A}$ and $K_{1B}$ mixing angle, where $K_{1A}$ and $K_{1B}$
    are the $^3 P_1 $ and  $^1 P_1 $ states of $K_1$, respectively, which are
     related to the physical $K_1 (1270)$ and $K_1 (1400)$ states.

We then study transitions of the heavy flavor pseudoscalar mesons to pseudoscalar
mesons ($P$), vector mesons ($V$), scalar mesons ($S$) and axial vector mesons ($A$)
within the CLF    model. Numerical results of the form factors for these
transitions and their momentum dependence are presented in
detail. In particular, all the form factors for heavy-to-light and
heavy-to-heavy transitions for charmed mesons $(D, D_s)$ and bottom mesons $(B, B_s)$
are calculated. Further, their sensitivity to uncertainties of $\beta$
parameters of the initial as well as of the
final mesons is investigated separately. Theoretically, the Isgur-Scora-Grinstein-Wise
(ISGW) quark model \cite{ISGW, IW89} has been the only model for a long time
 that could provide a systematical estimate of the transition
of a ground-state $s$-wave meson to a low-lying $p$-wave
meson. However, this model is based on the nonrelativistic constituent quark
picture. We have earlier pointed out
\cite{Cheng04} that relativistic effects could manifest
in heavy-to-light transitions at maximum recoil where the final-state meson can be
highly relativistic. For example, the
$B\rightarrow a_1$ form factor $V^{Ba_1}_0 (0)$ is found to be 0.13
in the relativistic light-front model \cite{Cheng04}, while it is as big as
1.01 in the ISGW model \cite{ISGW}.
Hence there is no reason to expect that the
nonrelativistic quark model is still applicable there, though in the
improved version of the model (ISGW2) \cite{ISGW2} a number of improvements,
such as the constraints imposed by heavy quark symmetry and
hyperfine distortions of wave functions have been incorporated.
We believe that the CLF quark model can provide
useful and reliable information on $B \rightarrow M$ transitions
particularly at maximum recoil.

The paper is organized as follows. The basic features of the covariant
light-front (CLF) model are recapitulated in Sec. II. In Sec. III, decay constants
are presented in the CLF model. Available experimental
measurements for various decays are used to determine decay constants,
which in turn are used to fix $\beta$ parameters of the CLF model.
Sometimes, lattice predictions for few decay constants are also used for this purpose.
  In Sec. IV, the analysis of form factors appearing for transitions from pseudoscalar mesons
  to $s$-wave mesons (pseudoscalar or vector) and $p$-wave mesons
   (scalar and axial vector)  is given. In Sec. V, numerical results are presented
   for these form factors and their $q^2$ - dependence taking proper
    inclusions of uncertainties in the shape parameter, $\beta$. Summary and conclusions
    are given in Sec. VI.

\section{Formalism of a covariant light-front model}

\begin{figure}[t!]
\centerline{
            {\epsfxsize2.1 in \epsffile{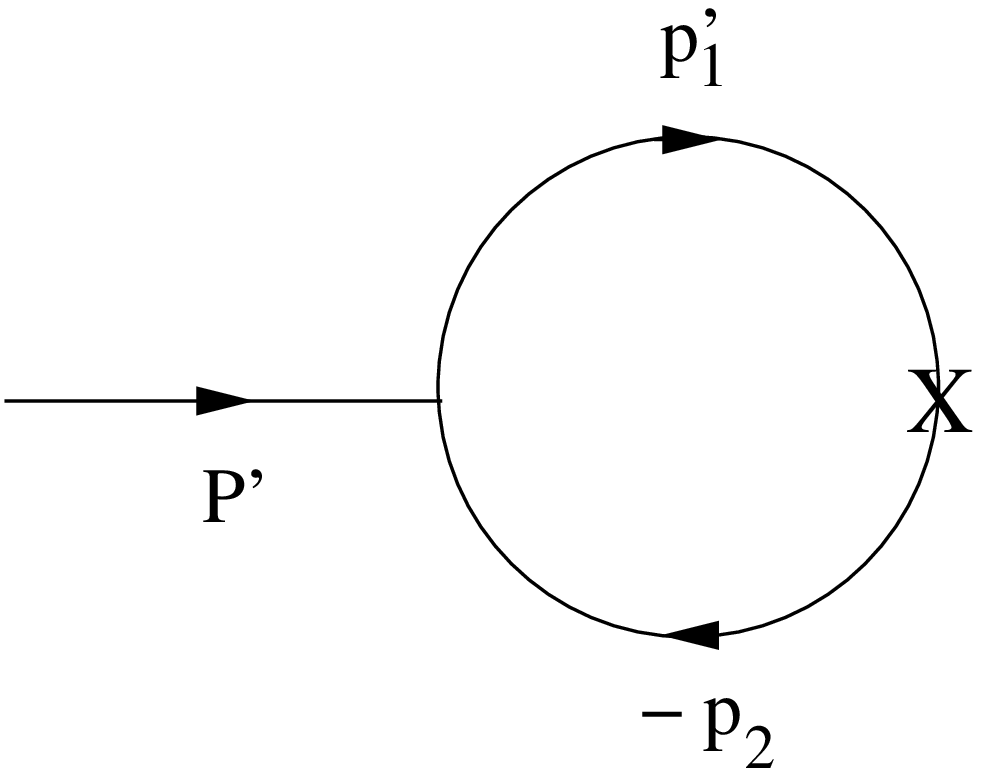}}
            \hspace{1cm}
            {\epsfxsize3 in \epsffile{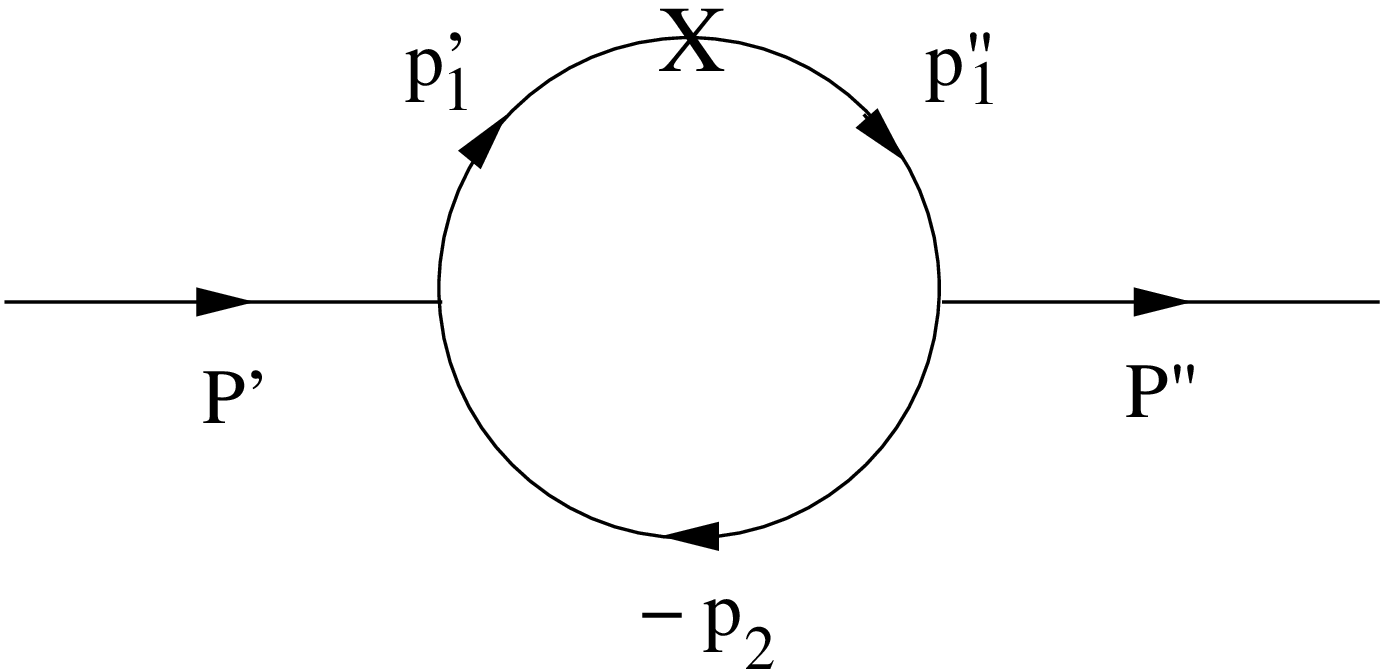}}}
\centerline{\,\,\,\,\,(a)\hspace{6.2cm}(b)} \vskip0.2cm
\caption{Feynman diagrams for (a) meson decay and (b) meson
transition amplitudes, where $P^{\prime(\pp)}$ is the incoming
(outgoing) meson momentum, $p^{\prime(\pp)}_1$ is the quark
momentum, $p_2$ is the anti-quark momentum and $X$ denotes the
corresponding $V-A$ current vertex.}\label{fig:feyn}
\end{figure}

In the conventional light-front framework, the constituent quarks
of the meson are required to be on their mass shells
and various physical quantities are
extracted from the plus component of the corresponding current
matrix elements. However, this procedure will miss the zero-mode
effects and render the matrix elements non-covariant. Jaus
\cite{Jaus90,Jaus91} has proposed a covariant light-front approach that
permits a systematical way of dealing with the zero mode
contributions. Physical quantities such as the decay constants and
form factors can be calculated in terms of Feynman momentum loop
integrals which are manifestly covariant. This of course means
that the constituent quarks of the bound state are off-shell. In
principle, this covariant approach will be useful if the vertex
functions can be determined by solving the QCD bound state
equation. In practice, we would have to be contended with the
phenomenological vertex functions such as those employed in the
conventional light-front model. Therefore, using the light-front
decomposition of the Feynman loop momentum, say $p_\mu$, and
integrating out the minus component of the loop momentum $p^-$,
one goes from the covariant calculation to the light-front one.
Moreover, the antiquark is forced to be on its mass shell after
$p^-$ integration. Consequently, one can replace the covariant
vertex functions by the phenomenological light-front ones.

To begin with, we consider decay and transition amplitudes given
by one-loop diagrams as shown in Fig.~\ref{fig:feyn} for the decay
constants and form factors of ground-state $s$-wave mesons and
low-lying $p$-wave mesons.  We follow the approach of
\cite{Jaus99, Cheng04} and use the same notation. The incoming (outgoing)
meson has the momentum $P^{\prime(\pp)}=p_1^{\prime(\pp)}+p_2$,
where $p_1^{\prime(\pp)}$ and $p_2$ are the momenta of the
off-shell quark and antiquark, respectively, with masses
$m_1^{\prime(\pp)}$ and $m_2$. These momenta can be expressed in
terms of the internal variables $(x_i, p_\bot^\prime)$,
 \be
 p_{1,2}^{\prime+}=x_{1,2} P^{\prime +},\qquad
 p^\prime_{1,2\bot}=x_{1,2} P^\prime_\bot\pm p^\prime_\bot,
 \en
with $x_1+x_2=1$. Note that we use $P^{\prime}=(P^{\prime +},
P^{\prime -}, P^\prime_\bot)$, where $P^{\prime\pm}=P^{\prime0}\pm
P^{\prime3}$, so that $P^{\prime 2}=P^{\prime +}P^{\prime
-}-P^{\prime 2}_\bot$.

In the covariant light-front approach, total four momentum is
conserved at each vertex where quarks and antiquarks are
off-shell. These differ from the conventional light-front approach
(see, for example ~\cite{Jaus91, Cheng97}) where the plus and
transverse components of momentum are conserved, and quarks as
well as antiquarks are on-shell.

It is useful to define some internal quantities for on-shell quarks:

\begin{eqnarray} \label{eq:internalQ}
 M^{\prime2}_0
          &=&(e^\prime_1+e_2)^2=\frac{p^{\prime2}_\bot+m_1^{\prime2}}
                {x_1}+\frac{p^{\prime2}_{\bot}+m_2^2}{x_2},\quad\quad
                \widetilde M^\prime_0=\sqrt{M_0^{\prime2}-(m^\prime_1-m_2)^2},
 \nonumber\\
 e^{(\prime)}_i
          &=&\sqrt{m^{(\prime)2}_i+p^{\prime2}_\bot+p^{\prime2}_z},\quad\qquad
 p^\prime_z=\frac{x_2 M^\prime_0}{2}-\frac{m_2^2+p^{\prime2}_\bot}{2 x_2 M^\prime_0}.
 \end{eqnarray}
Here $M^{\prime2}_0$ can be interpreted as the kinetic invariant
mass squared of the incoming $q\bar q$ system, and $e_i$ the
energy of the quark $i$.

\begin{table}[b]
\caption{\label{tab:feyn} Feynman rules for the vertices
($i\Gamma^\prime_M$) of the incoming mesons-quark-antiquark, where
$p^\prime_1$ and $p_2$ are the quark and antiquark momenta,
respectively. Under the contour integrals to be discussed below,
$H^\prime_M$ and $W^\prime_M$ are reduced to $h^\prime_M$ and
$w^\prime_M$, respectively, whose expressions are given by
Eq.~(\ref{eq:h}). Note that for outgoing mesons, we shall use
$i(\gamma_0\Gamma^{\prime\dagger}_M\gamma_0)$ for the
corresponding vertices.}
\begin{tabular}{|c| c|}
\hline
 $M\,(^{2S+1}L_J) $
      &$i\Gamma^\prime_M$
      \\
      \hline
 pseudoscalar ($^1S_0$)
      &$H^\prime_P\gamma_5$
      \\
 vector ($^3S_1$)
      &$i H^\prime_V [\gamma_\mu-\frac{1}{W^\prime_V}(p^\prime_1-p_2)_\mu]$
      \\
 scalar ($^3P_0$)
      &$-i H^\prime_S$
      \\
 axial ($^3 P_1$)
      &$-i H^\prime_{^3\!A}[\gamma_\mu+\frac{1}{W^\prime_{^3\!A}}(p^\prime_1-p_2)_\mu]\gamma_5$
      \\
 axial ($^1 P_1$)
      &$-i H^\prime_{^1\!A} [\frac{1}{W^\prime_{^1\!A}}(p^\prime_1-p_2)_\mu]\gamma_5$\\
\hline
\end{tabular}
\end{table}

It has been shown in \cite{CM69} that one can pass to the
light-front approach by integrating out the $p^-$ component of the
internal momentum in covariant Feynman momentum loop integrals.
We need Feynman rules for the meson-quark-antiquark vertices to
calculate the amplitudes shown in Fig.~1. These Feynman rules for
vertices ($i\Gamma^\prime_M$) of ground-state $s$-wave mesons and
low-lying $p$-wave mesons are summarized in Table~\ref{tab:feyn}.
Next, we shall find the decay constants in the covariant light-front approach.

\section{Decay constants}

The decay constants for $J=0,1$ mesons are defined by the matrix
elements
 \be \label{eq:AM}
  \la 0|A_\mu|P(P^\prime)\ra &\equiv& {\cal A}^{P}_\mu=i  f_P P^\prime_\mu ,\qquad
      \la 0|V_\mu|S(P^\prime)\ra\equiv {\cal A}^{S}_\mu= f_S P^\prime_\mu ,
  \\
  \la 0|V_\mu|V(P^\prime,\vp')\ra &\equiv& {\cal A}^{V}_\mu=M^\prime_V
  f_V\vp^\prime_\mu,\quad
      \la 0|A_\mu|\,^{3(1)}\!A(P^\prime,\vp')\ra \equiv {\cal A}^{^3\!A(^1\!A)}_\mu
      =M^\prime_{^3\!A(^1\!A)} f_{^3\!A(^1\!A)}\varepsilon^\prime_\mu, \non
\en
where the $^{2S+1} L_J= {}^1S_0$, $^3P_0$, $^3S_1$, $^3P_1$ and
$^1P_1$ states of $q_1^\prime \bar q_2$ mesons are
denoted by $P$, $S$, $V$, $^3\!A$ and $^1\!A$, respectively.
It is
useful to note that in the SU(N)-flavor limit ($m_1^\prime=m_2$)
we should have vanishing $f_S$ and $f_{^1\!A}$. The former can be
seen by applying equations of motion to the matrix element of the
scalar resonance in Eq. (\ref{eq:AM}) to obtain
 \be \label{eq:Seom}
 m_S^2f_S=\,i(m'_1-m_2)\la 0|\bar q_1q_2|S\ra.
 \en
The latter is based on the argument that the light $^3P_1$ and
$^1P_1$ states transfer under charge conjugation as
 \be
 M_a^b(^3P_1) \to M_b^a(^3P_1), \qquad M_a^b(^1P_1) \to
 -M_b^a(^1P_1),~~~(a=1,2,3),
 \en
where the light axial-vector mesons are represented by a $3\times
3$ matrix. Since the weak axial-vector current transfers as
$(A_\mu)_a^b\to (A_\mu)_b^a$ under charge conjugation, it is clear
that the decay constant of the $^1P_1$ meson vanishes in the SU(3)
limit \cite{Suzuki}. This argument can be generalized to heavy
axial-vector mesons. In fact, under similar charge conjugation
argument [$(V_\mu)_a^b\to -(V_\mu)_b^a$, $M_a^b(^3P_0) \to
M_b^a(^3P_0)$] one can also prove the vanishing of $f_S$ in the
SU(N) limit.

Furthermore, in the heavy quark limit ($m_1^\prime\to\infty$), the
heavy quark spin $s_Q$ decouples from the other degrees of freedom
so that $s_Q$ and the total angular momentum of the light
antiquark $j$ are separately good quantum numbers. Hence, it is
more convenient to use the $L^j_J=P^{3/2}_2$, $P^{3/2}_1$,
$P^{1/2}_1$ and $P^{1/2}_0$ basis. It is obvious that the first
and the last of these states are $^3P_2$ and $^3P_0$,
respectively, while \cite{IW91}
\begin{equation} \label{eq:Phalf}
\left|P^{3/2}_1\right\ra=\sqrt{\frac{2}{3}}\,\left|^1P_1\right\ra
+{1\over \sqrt{3}}\,\left|^3P_1\right\ra,\qquad
\left|P^{1/2}_1\right\ra={1\over \sqrt{3}}\,\left|^1P_1\right\ra
-\sqrt{\frac{2}{3}}\,\left|^3P_1\right\ra.
\end{equation}
Heavy quark symmetry (HQS) requires~\cite{IW89,HQfrules}
\begin{equation} \label{eq:HQSf}
 f_V=f_P,\qquad
 f_{A^{1/2}}=f_S,\qquad
 f_{A^{3/2}}=0,
\end{equation}
where we have denoted the $P^{1/2}_1$ and $P^{3/2}_1$ states by
$A^{1/2}$ and $A^{3/2}$, respectively.
These relations in the above equation can be understood from the
fact that $(S^{1/2}_0,S^{1/2}_1)$, $(P_0^{1/2},P_1^{1/2})$ and
$(P_1^{3/2},P_2^{3/2})$ form three doublets in the HQ limit and
that the tensor meson cannot be induced from the $V-A$ current.

Following the procedure described in \cite{Jaus99, Cheng04},
we now evaluate meson decay constants through the following formulas:
 \be
 f_P=\frac{N_c}{16\pi^3}\int dx_2 d^2p^\prime_\bot \frac{h^\prime_P}{x_1
 x_2 (M^{\prime2}-M^{\prime2}_0)}4(m_1^\prime x_2+m_2 x_1),
 \label{eq:fP}
 \en

 \be
 f_V&=&\frac{N_c}{4\pi^3M^\prime}\int dx_2 d^2p^\prime_\bot \frac{h^\prime_V}{x_1
      x_2 (M^{\prime2}-M^{\prime2}_0)}
 \nonumber\\
      &&\qquad\qquad\times\left[x_1 M^{\prime2}_0-m_1^\prime (m_1^\prime-m_2)-p^{\prime2}_\bot
            +\frac{m_1^\prime+m_2}{w^\prime_V}\,p^{\prime2}_\bot \right],
 \label{eq:fV}
 \en

\be
 f_S=\frac{N_c}{16\pi^3}\int dx_2 d^2p^\prime_\bot \frac{h^\prime_S}{x_1
 x_2 (M^{\prime2}-M^{\prime2}_0)}4(m_1^\prime x_2-m_2 x_1),
 \label{eq:fS}
 \en

  \be
 f_{^3\!A}&=&-\frac{N_c}{4\pi^3M^\prime}\int dx_2 d^2p^\prime_\bot
            \frac{h^\prime_{^3\!A}}{x_1 x_2 (M^{\prime2}-M^{\prime2}_0)}
 \nonumber\\
      &&\qquad\qquad\times\left[x_1 M^{\prime2}_0-m_1^\prime (m_1^\prime+m_2)-p^{\prime2}_\bot
            -\frac{m_1^\prime-m_2}{w^\prime_{^3\!A}}\,p^{\prime2}_\bot \right],
 \nonumber\\
  f_{^1\!A}&=&\frac{N_c}{4\pi^3 M^\prime}\int dx_2 d^2p^\prime_\bot
            \frac{h^\prime_{^1\!A}}{x_1 x_2 (M^{\prime2}-M^{\prime2}_0)}
      \left(\frac{m_1^\prime-m_2}{w^\prime_{^1\!A}}\,p^{\prime2}_\bot \right),
 \label{eq:fA}
 \en

where

\begin{eqnarray} \label{eq:vertex}
 h^\prime_P&=&h^\prime_V
                  =(M^{\prime2}-M_0^{\prime2})\sqrt{\frac{x_1 x_2}{N_c}}
                    \frac{1}{\sqrt{2}\widetilde M^\prime_0}\varphi^\prime,
 \nonumber\\
 h^\prime_S &=&\sqrt{\frac{2}{3}}h^\prime_{^3\!A}
                  =(M^{\prime2}-M_0^{\prime2})\sqrt{\frac{x_1 x_2}{N_c}}
                    \frac{1}{\sqrt{2}\widetilde M^\prime_0}\frac{\widetilde
                     M^{\prime
                     2}_0}{2\sqrt{3}M^\prime_0}\varphi^\prime_p,
       \nonumber\\
 h^\prime_{^1\!A}&=& h^\prime_T =(M^{2\prime}-M_0^{\prime 2})\sqrt{\frac{x_1
 x_2}{N_c}}\frac{1}{\sqrt{2}\widetilde M^\prime_0}\varphi'_p\, ,
 \non\\
 w^\prime_V&=&M^\prime_0+m^\prime_1+m_2,\quad
 w^\prime_{^3\!A}=\frac{\widetilde{M}'^2_0}{m^\prime_1-m_2},\quad
 w^\prime_{^1\!A}=2\,,
 \label{eq:h}
\end{eqnarray}
are the appropriate replacements of the vertex functions,
 \be
H^\prime_M
       &\to&\hat H^\prime_M
           =H^\prime_M(\hat p^{\prime 2}_1,\hat p^2_2)
           \equiv h^\prime_M,
\non\\
W^\prime_M
       &\to&\hat W^\prime_M
           =W^\prime_M(\hat p^{\prime 2}_1,\hat p^2_2)
           \equiv w^\prime_M,
 \label{eq:contourA}
 \en
appearing in the matrix elements of annihilation of a meson state via weak currents,
and $\varphi'$ and $\varphi'_p$ are the light-front momentum
distribution amplitudes for $s$-wave and $p$-wave mesons,
respectively. There are several popular phenomenological
light-front wave functions that have been employed to describe
various hadronic structures in the literature. In the present
work, we shall use the Gaussian-type wave function \cite{Gauss}
\begin{eqnarray} \label{eq:Gauss}
 \varphi^\prime
    &=&\varphi^\prime(x_2,p^\prime_\perp)
             =4 \left({\pi\over{\beta^{\prime2}}}\right)^{3\over{4}}
               \sqrt{{dp^\prime_z\over{dx_2}}}~{\rm exp}
               \left(-{p^{\prime2}_z+p^{\prime2}_\bot\over{2 \beta^{\prime2}}}\right),
\nonumber\\
 \varphi^\prime_p
    &=&\varphi^\prime_p(x_2,p^\prime_\perp)=\sqrt{2\over{\beta^{\prime2}}}~\varphi^\prime,\quad\qquad
         \frac{dp^\prime_z}{dx_2}=\frac{e^\prime_1 e_2}{x_1 x_2 M^\prime_0}.
 \label{eq:wavefn}
\end{eqnarray}
The parameter $\beta'$, which describes the momentum distribution,
is expected to be of order $\Lambda_{\rm QCD}$.

Note that with the explicit form of
$h'_P$ shown in Eq.~(\ref{eq:h}), the familiar expression of $f_P$
in the conventional light-front approach~\cite{Jaus91,Cheng97},
namely,
 \be \label{eq:fP0}
  f_P=2\frac{\sqrt{2N_c}}{16\pi^3}\int dx_2 d^2p^\prime_\bot \frac{1}{\sqrt{x_1
 x_2} \widetilde M'_0}\,(m_1^\prime x_2+m_2
 x_1)\,\varphi^\prime(x_2,p^\prime_\perp),
 \en
is reproduced. For decay constants of vector and axial-vector
mesons, we consider the case with the transverse polarization given by
 \be
 \vp(\pm)=\left(\frac{2}{P^{\prime+}}\varepsilon_\bot\cdot
 P^\prime_\bot,0,\varepsilon_\bot\right),\qquad
 \varepsilon_\bot=\mp\frac{1}{\sqrt2}(1,\pm i).
 \en
For $m_1^\prime=m_2$, the meson wave function is symmetric with
respect to $x_1$ and $x_2$, and hence $f_S=0$, as it should be.
Similarly, it  is clear that $f_{^1\!A}=0$ for $m^\prime_1=m_2$. The
SU(N)-flavor constraints on $f_S$ and $f_{^1\!A}$ are thus
satisfied.\footnote{We wish to stress that the vector decay constant obtained in the
conventional light-front model \cite{Jaus91} does not coincide
with the above result (\ref{eq:fV}) owing to the missing zero mode
contribution.}

\begin{table}[b!]
\caption{\label{tab:beta} The input parameter $\beta$ (in units of
GeV) in the Gaussian-type wave function (\ref{eq:wavefn}) for mesons. Note that
$\beta_{q\bar q}$ is used for the $(u\overline{u} + d\overline{d})/\sqrt{2}$ state.}
\begin{ruledtabular}
\begin{tabular}{|c|ccccc|}
$^{2S+1} L_J$
          & $^1S_0$
          & $^3S_1$
          & $^3P_0$
          & $^3P_1$
          & $^1P_1$
          \\
\hline
$\beta_{d\bar u}$
          & $0.3077^{+0.0009}_{-0.0008}$
          & $0.2815^{+0.0046}_{-0.0047}$
          & $0.2983^{+0.0123}_{-0.0129}$
          & $0.2983^{+0.0123}_{-0.0129}$
          & $0.2983^{+0.0123}_{-0.0129}$
          \\
$\beta_{q\bar q}$
          & $0.3499^{+0.0136}_{-0.0129}$
          & $0.2640^{+0.0031}_{-0.0032}$
          & $0.2983\pm 0.0298$
          & $0.2983\pm 0.0298$
          & $0.2983\pm 0.0298$
          \\
$\beta_{s\bar u}$
          & $0.3479^{+0.0029}_{-0.0029}$
          & $0.2926^{+0.0047}_{-0.0047}$
          & $0.3224^{+0.0163}_{-0.0195}$
          & $0.3224^{+0.0163}_{-0.0195}$
          & $0.3224^{+0.0163}_{-0.0195}$
          \\
$\beta_{s\bar s}$
          & $0.3598^{+0.0220}_{-0.0208}$
          & $0.3083\pm 0.0014$
          & $0.3492\pm 0.0064$
          & $0.3492\pm 0.0064$
          & $0.3492\pm 0.0064$
          \\
\hline
$\beta_{c\bar u}$
          & $0.4656^{+0.0217}_{-0.0212}$
          & $0.4255\pm 0.0426$
          & $0.3890\pm 0.0389$
          & $0.3890\pm 0.0389$
          & $0.3890\pm 0.0389$
          \\
$\beta_{c\bar s}$
          & $0.5358^{+0.0137}_{-0.0135}$
          & $0.4484\pm 0.0448$
          & $0.3900\pm 0.0390$
          & $0.3900\pm 0.0390$
          & $0.3900\pm 0.0390$
          \\
$\beta_{c\bar c}$
          & $0.7690^{+0.0049}_{-0.0049}$
          & $0.6492\pm 0.0069$
          & $0.4200\pm 0.0420$
          & $0.4200\pm 0.0420$
          & $0.4200\pm 0.0420$
          \\
\hline
$\beta_{b\bar u}$
          & $0.5547^{+0.0260}_{-0.0261}$
          & $0.5183\pm 0.0518$
          & $0.5000\pm 0.0500$
          & $0.5000\pm 0.0500$
          & $0.5000\pm 0.0500$
          \\
$\beta_{b\bar s}$
          & $0.6103^{+0.0330}_{-0.0331}$
          & $0.5589\pm 0.0559$
          & $0.5500\pm 0.0550$
          & $0.5500\pm 0.0550$
          & $0.5500\pm 0.0550$
          \\
$\beta_{b\bar c}$
          & $0.9582\pm 0.0958$
          & $0.8451\pm 0.0845$
          & $0.6800\pm 0.0680$
          & $0.6800\pm 0.0680$
          & $0.6800\pm 0.0680$
          \\
$\beta_{b\bar b}$
          & $1.4514\pm 0.0132$
          & $1.3267\pm 0.0100$
          & $0.9993\pm 0.0999$
          & $0.9993\pm 0.0999$
          & $0.9993\pm 0.0999$
          \\
\end{tabular}
\end{ruledtabular}
\end{table}

To perform numerical computations of decay constants and form factors, we need to
specify the input parameters in the covariant light front model.
These are the constituent quark masses and the shape parameter $\beta$
appearing in the Gaussian-type wave function (\ref{eq:wavefn}).
For constituent quark masses, we
use~\cite{Jaus96,Cheng97,Hwang02,Jaus99, Cheng04}
 \be \label{eq:quarkmass}
m_{u,d}=0.26\,{\rm GeV},\qquad m_s=0.45\,{\rm GeV},\qquad
m_c=1.40\,{\rm GeV},\qquad m_b=4.64\,{\rm GeV}.
 \en

Shown in Tables~\ref{tab:beta} and \ref{tab:f} are the input parameter
$\beta$ and decay constants, respectively. In Table \ref{tab:f}
the decay constants in parentheses are used to determine $\beta$ using the
analytic expressions in the covariant light-front model as given above.
For most of $s$-wave mesons, and a few axial vector mesons, these are fixed from
the latest decay rates given in the Particle Data Group ~\cite{PDG10},
or other analysis based on some experimental results. For decay constants of
some heavy flavor mesons, we have used recent lattice results to
fix $\beta$.  For the remaining $p$-wave mesons,  we use the $\beta$ parameters
 obtained in the ISGW2 model~\cite{ISGW2}, the improved version of the ISGW
model, up to some simple scaling. In this paper, we have investigated
the variation of the form factors and their slope parameters for $q^2$
 dependence with the variation of $\beta$ values.
 Wherever the experimental information is available, we have
used that to fix the errors for the corresponding $\beta$ values,
otherwise arbitrarily introduced an uncertainty of $10 \%$  in $\beta$
for some $s$-wave and $p$-wave mesons.

\begin{table}[t!]
\caption{\label{tab:f} Meson decay constants (in units of MeV)
obtained by using Eqs.~(\ref{eq:fP}), (\ref{eq:fS}), (\ref{eq:fV})
and (\ref{eq:fA}). Those in parentheses are taken as inputs to
determine the corresponding $\beta$'s shown in
Table~\ref{tab:beta}. Decay constants of some $p$-wave mesons are
also used as inputs (see the text for details). Here
$f_{q\bar q}$ denotes decay constant for the
$(u\overline{u} + d\overline{d})/\sqrt{2}$ state.}
\begin{ruledtabular}
\begin{tabular}{|c|ccccc|}
$^{2S+1} L_J$
          & $^1S_0$
          & $^3S_1$
          & $^3P_0$
          & $^3P_1$
          & $^1P_1$
          \\
\hline
$f_{d\bar u}$
          & $(130.41\pm 0.20)$
          & $(215\pm 5)$
          & $0$
          & $(-203\mp 18)$
          & $0$
          \\
$f_{q\bar q}$
          & $(139.54\pm 2.62)$
          & $(195\pm 3)$
          & $0$
          & $-193^{-43}_{+38}$
          & $0$
          \\
$f_{s\bar u}$
          & $(156.1\pm 0.9)$
          & $(217\pm 5)$
          & $34.9^{+1.4}_{-1.8}$
          & $(-212^{-23}_{+26})$
          & $20.4^{+1.5}_{-1.8}$
          \\
$f_{s\bar s}$
          & $(174.75\pm7.83)$
          & $(228\pm 2)$
          & $0$
          & $(-230\mp9)$
          & $0$
          \\
\hline
$f_{c\bar u}$
          & $(206.7\pm8.9)$
          & $(245)^{+35}_{-34}$
          & $107\pm13$
          & $-177^{-38}_{+34}$
          & $59.6^{+9.8}_{-9.5}$
          \\
$f_{c\bar s}$
          & $(254.6\pm5.9)$
          & $(272)^{+39}_{-38}$
          & $74.4^{+10.4}_{-10.6}$
          & $-159^{-36}_{+32}$
          & $42.2^{+7.6}_{-7.3}$
          \\
$f_{c\bar c}$
          & $(394.7\pm2.4)$
          & $(411\pm 6)$
          & $0$
          & $-105^{-26}_{+23}$
          & $0$
          \\
\hline
$f_{b\bar u}$
          & $(193\pm11)$
          & $(196)^{+28}_{-27}$
          & $143\pm21$
          & $-155^{-30}_{+28}$
          & $83.6^{+14.3}_{-13.6}$
          \\
$f_{b\bar s}$
          & $(231\pm15)$
          & $(229)^{+32}_{-31}$
          & $139\pm22$
          & $-166^{-34}_{+31}$
          & $82.6^{+15.0}_{-14.2}$
          \\
$f_{b\bar c}$
          & $440^{+51}_{-52}$
          & $440^{+51}_{-52}$
          & $90.6^{+17.5}_{-16.6}$
          & $-155^{-37}_{+33}$
          & $52.0^{+11.2}_{-10.3}$
          \\
$f_{b\bar b}$
          & $(708\pm8)$
          & $(708\pm8)$
          & $0$
          & $-185^{-48}_{+42}$
          & $0$
          \\
\end{tabular}
\end{ruledtabular}
\end{table}

Several remarks are in order:

(i) Decay constants of the charged pseudoscalar mesons,  $\pi^+, K^+, D^+, D^+_s,$ and $B^-$
(and their charge-conjugate partners) can be determined from their purely leptonic
 decay rates. These mesons formed from a quark and anti-quark can decay to a charged lepton pair
  when their constituents annihilate via a virtual $W$ boson. Now quite precise measurements
   are available for the branching fractions of $P \to \ell \nu_{\ell}$ decays \cite{PDG10}. Following the
   analysis of Rosner and Stone \cite{Rosner10} for the available
   branching fractions,  we take
    $f_{\pi}= 130.41\pm 0.20, f_K = 156.10\pm0.85, f_D = 206.7\pm8.9$
    (all in MeV) to fix the $\beta$ parameters of the respective mesons.

(ii) For fixing $\beta_{D_s}$, we have taken the world average value $254.6\pm5.9$ MeV
 for $f_{D_s}$ given by the Heavy Flavor Averaging Group \cite{HFAG10} based on
  the BaBar, Belle and CLEO measurements of ${\cal B}( D^+_s \rightarrow \mu^+\nu)$
  and ${\cal B}(D^+_s \rightarrow \tau^+\nu)$. This value can be compared well to the
   results from the two precise lattice QCD calculations
   $f_{D_s} = 248.0 \pm 2.5 $ MeV and $249 \pm 11$ MeV, respectively, from the HPQCD
   Collaboration \cite{HPQCD10} and the Fermilab/MILC Collaboration  \cite{MILC08}.
    For the bottom sector, the Belle and BaBar collaborations have found
    evidence for $B^- \rightarrow \tau^- \nu$ decay in $e^+ e^-\rightarrow
B^-B^+$ collisions at the $\Upsilon(4S)$ energy, however, the errors are rather
 large in the measured branching fractions with the computed average value
  ${\cal B}(B^- \rightarrow \tau^- \nu) = 1.72^{+0.43}_{-0.42} \times 10^{-4}$. Further a more
   accurate value of $|V_{ub}|$ is required for the determination of $f_B$.
   Considering the large uncertainties on $V_{ub}$ and the branching fraction
   measurements for $B^- \rightarrow \tau^- \nu$,  and sensitivity of this decay
    to the new physics, we rely upon  $f_B = 193 \pm 11 $ MeV, used in \cite{Rosner10} as the
    average of the two lattice results $f_B = 195 \pm 11$ MeV \cite{MILC08} and
      $f_B = 190 \pm 13 $ MeV \cite{HPQCD09}, to fix the input parameter $\beta_{B}$.
       Likewise, for $B_s$ meson, we use the lattice prediction of $f_{B_s} = 231 \pm 15$ MeV
        \cite{HPQCD09} for determining $\beta_{B_s}$.

(iii) The decay constants of the diagonal pseudoscalar mesons
$\pi^0, \eta, \eta^\prime$ and $\eta_c$,
 in principle, could be obtained from $P \rightarrow \gamma \gamma$ branching fractions.
  In the case of $\pi^0$, the value of $f_{\pi^0}= 130 \pm 5$ MeV \cite{Suzuki03} has been extracted
  from the measured $\pi^0 \rightarrow \gamma \gamma$ decay width, which is compatible
   with $f_{\pi^{\pm}}$, as is expected from isospin symmetry. However, decay constants
   of the $\eta -\eta^\prime$ system cannot be extracted from two-photon decay rates
   alone and get more complicated due to the $\eta -\eta^\prime$ mixing, the chiral
    anomaly and gluonium mixing \cite{Feldmann00, ChengLi09}. For describing the
     mixing between $\eta$ and $\eta^\prime$, it is more convenient to employ
     the flavor states $(u\overline{u} + d\overline{d})/\surd {2}$, and
$ (s\overline{s})$ labeled by the $\eta_q$ and $\eta_s$, respectively. We then write
 \be \label{eq:EtaMixing}
\eta = \eta_q \cos\phi - \eta_s \sin\phi,
 \nonumber\\
 \eta^\prime = \eta_q \sin\phi +  \eta_s\cos\phi,
 \en
 where $\phi = (39.3 \pm 1.0)^ \circ$ follows from the analysis of Feldmann {\it et al.} \cite{Feldmann00}
 to fit the experimental data. This analysis also gives
 $f_{\eta}/f_{\pi}= 1.07 \pm 0.02$ and  $f_{\eta^{\prime}}/f_{\pi}= 1.34 \pm 0.06$, which are used
 in the present work.
      For $\eta_c$, the decay width is poorly known with PDG \cite{PDG10} estimate
       given as $\Gamma (\eta_c \rightarrow \gamma \gamma)=7.2\pm2.1$ keV
       giving $f_{\eta_c} = 0.4 \pm 0.1 $ GeV. Alternatively,
       one may extract $f_{\eta_c}$ from $B \rightarrow \eta_c K $ decay using the
       factorization approximation, for which CLEO \cite{Edwards01} obtained $f_{\eta_c} = 335 \pm 75$ MeV.
       In the literature, $f_{\eta_c}$  is expected to be quite close to $f_{J/\psi}$
         on the basis of quark model considerations \cite{Gourdin95}. Recently, the
        HPQCD collaboration \cite{HPQCD10} has reported a more precise result for
         $f_{\eta_c}$ to be  $394.7\pm2.4$ MeV consistent with other estimates,
         and is in fact very close to the experimental result $f_{J/\psi}= 410.6\pm 6.2$ MeV
         obtained from the leptonic decay width of $J/\psi$ \cite{PDG10}. So we use the
         lattice prediction to fix $\beta_{\eta_c}$. In the absence of
          any experimental estimate for $f_{\eta_b}$, we shall assume
           $f_{\eta_b} \approx f_{\Upsilon}$ to fix $\beta_{\eta_b}$ following the heavy-quark
           spin symmetry.

(iv) For vector mesons, we extract the decay constants for diagonal states from
the experimental values of their respective branching fractions of leptonic decays
 $V \rightarrow l^+ l^-$ decays \cite{PDG10}. Thus we obtain
  $f_{\rho^0}= 221.20 \pm 0.94,~ f_{\omega} = 194.60 \pm 3.24,~
  f_{\phi}= 227.9 \pm 1.5,~ f_{J/\psi}= 410.6\pm 6.2$ and $f_{\Upsilon} = 708.0\pm7.8$ (all in MeV)
  for ideal mixing, and use them to fix  the $\beta_V$ parameters of the respective mesons.

(v) The decay constant $f_V$ determines not only the coupling of the neutral
vector mesons to a photon, but also the coupling of charged vector mesons, like
$\rho^{\pm}$ and $K^{*\pm}$,
to the weak vector bosons $W^{\pm}$. There are no data available for the leptonic
 decay of these charged vector mesons, but the couplings can be extracted indirectly
from the decays $\tau \rightarrow \rho \nu_{\tau}$ and
 $\tau \rightarrow K^* \nu_{\tau}$. With the experimental values for the
  branching fractions of these decays ${\cal B}(\tau \rightarrow \rho \nu_{\tau})= 25.02 \% $
  and ${\cal B}(\tau \rightarrow K^* \nu_{\tau})= 1.28  \% $, the decay width formula
 \be \label{eq:tauVdec}
  \Gamma(\tau \rightarrow V \nu_{\tau}) = {{G^2_F} \over {16 \pi}} |V_{q_1 q_2}|^2 f^2_{V} {{(m^2_{\tau}+2 m^2_V)(m^2_{\tau}- m^2_V)^2} \over {m^3_{\tau}}},
 \en
where $V_{q_1 q_2}$  is the appropriate CKM- factor corresponding to the vector meson $V$,
   yield $f_{\rho^{\pm}} = 209\pm4$ MeV and  $f_{K^{* \pm}} = 217 \pm 5$ MeV, respectively.
It is worth noting that the difference in $f_{\rho^0}$ and $f_{\rho^{\pm}}$
seems consistent with the expected size of isospin breaking, and we take the
average of the two values, i.e., $f_{\rho}= 215\pm 5$ MeV, the error chosen
so as to satisfy the two cases in extreme limits.

(vi) Contrary to the non-strange charmed
meson case, where $D^*$ has a slightly larger decay constant than
$D$, the recent measurements of $B\to D_s^{(*)}D^{(*)}$
\cite{PDG10,BaBarDs} indicate that the decay constants of $D_s^*$
and $D_s$ are relatively similar. As for the decay constant of $B^*$, a recent lattice calculation
yields $f_{B^*}/f_B=1.01\pm0.01^{+0.04}_{-0.01}$ \cite{Bernard}.
Explicitly, for naked charmed and bottom states $D^*, D^*_s, B^*$, and $B^*_s$,
 we have used the lattice predictions, $f_{D^*}= 245, f_{D^*_s}= 272, f_{B^*}= 196$, and $f_{B^*_s} = 229$
  (all in MeV) \cite{Becirevic99} to fix the central value of the respective parameters $\beta$,
  and allow $10\%$ variation in each case, giving decay constant ratios as
  $f_{D^*}/f_D = 1.18 \pm 0.17, f_{D^*_s}/f_{D_s} = 1.07 \pm 0.15$, and
  $ f_{B^*}/f_B \approx f_{B^*_s}/f_{B_s}  \approx 1.0 \pm 0.15 $, to leave the
  scope for matching with other results.

(vii) For axial vector mesons, there are two different nonets of
 $J^P = 1^+$ in the quark model as the orbital excitation of the
 $q \overline{q}$ system. In terms of the spectroscopic notation
  $^{2S+1} L_J$, there are two types of $p$-wave axial vector mesons,
  namely, $^3 P_1$ and $^1 P_1$, which have distinctive C quantum
numbers, $C=+$ and $C=-$, respectively. Experimentally,
the $J^{PC}= 1^{++}$ nonet consists of $a_1 (1260),
f_1 (1285), f_1 (1420)$, and $K_{1A}$, while the $J^{PC}= 1^{+-}$  nonet has
$b_1 (1235), h_1 (1170), h_1 (1380)$, and $K_{1B}$.

(viii)  It is generally argued that $a_1(1260)$ should have a
similar decay constant as the $\rho$ meson.
Presumably, $f_{a_1}$ can be extracted from the decay
$\tau\to a_1(1260)\nu_\tau$. Though this decay is not shown in the
Particle Data Group \cite{PDG10}, an experimental value of
$|f_{a_1}|=203\pm 18$ MeV is nevertheless quoted in
\cite{Bloch}.\footnote{The decay constant of $a_1$ can be tested
in the decay $B^+\to \bar D^0a_1^+$ which receives the main
contribution from the color-allowed amplitude proportional to
$f_{a_1}F^{BD}(m_{a_1}^2)$.} The $a_1 (1260)$ decay constant
$f_{a_1} = 238 \pm 10$ MeV obtained using the QCD sum rule method
 \cite{Yang07} is slightly higher than this value as well
 as $f_{\rho}= 215$ MeV. In Table \ref{tab:f} we have employed $f_{a_1}=-203\pm 18$ MeV
as input following our sign convention.

(ix) The nonstrange axial-vector mesons, for example, $a_1 (1260)$
and $b_1 (1235)$ cannot have mixing because
of the opposite $C$-parities. On the contrary, physical strange
 axial-vector mesons are the mixture of $^3P_1$ and $^1P_1$ states, while
  the heavy axial-vector resonances are generally taken as the mixture of $P_1^{1/2}$ and
$P_1^{3/2}$. For example, the physical
mass eigenstates $K_1 (1270)$ and $K_1 (1400)$ are a mixture of
$K_{1A}$ and $K_{1B}$ states owing to the mass difference of the
strange and nonstrange light quarks:
 \be \label{eq:K1mixing}
 K_1(1270)=K_{1A} \sin\theta_{K_1} +K_{1B}\cos\theta_{K_1},
 \nonumber\\
 K_1(1400)=K_{1A} \cos\theta_{K_1}-K_{1B}\sin\theta_{K_1}.
 \en
Using the experimental results ${\cal B}(\tau \rightarrow K_1 (1270) \nu_{\tau})=
(4.7 \pm 1.1) \times 10^{-3} $ and $\Gamma(\tau \rightarrow K_1 (1270) \nu_{\tau})/
[ \Gamma (\tau \rightarrow K_1 (1270) \nu_{\tau})+\Gamma (\tau \rightarrow K_1 (1400)
 \nu_{\tau})]= 0.69\pm 0.15$ \cite{PDG10}, and the decay width formula similar
 to that given in Eq. (\ref{eq:tauVdec}) with the replacement $V \to A$,
 we obtain \footnote{The large experimental error with the $K_1 (1400)$
  production in the $\tau$ decays, namely ${\cal B}(\tau \rightarrow K_1 (1400) \nu_{\tau})=
  (1.7 \pm 2.6) \times 10^{-3}$ \cite{PDG10}, does not provide sensible
   information for the $K_1 (1400)$ decay constant}
\be \label{eq:K1decayConst}
| f_{K_1(1270)}|= 169.5^{+18.8}_{-21.2}\, \textrm{MeV},
 \nonumber\\
 | f_{K_1(1400)}|= 139.2^{+41.3}_{-45.6}\, \textrm{MeV}.
 \en
These decay constants are related to $f_{K_{1A}}$ and $f_{K_{1B}}$ through
 \be \label{eq:K1decayConstMixing}
 m_{K_1 (1270)} f_{K_1 (1270)} =  m_{K_{1A}} f_{K_{1A}} \sin\theta_{K_1} + m_{K_{1B}} f_{K_{1B}} \cos\theta_{K_1},
  \nonumber\\
  m_{K_1 (1400)} f_{K_1 (1400)} =  m_{K_{1A}} f_{K_{1A}} \cos\theta_{K_1} - m_{K_{1B}} f_{K_{1B}} \sin\theta_{K_1},
 \en
where use of Eq. (\ref{eq:K1mixing}) and expressions for decay constants have been made.
From the analytic expressions of decay constants given in Eq. (\ref{eq:AM}),
it is clear that  $m_{K_{1A}} f_{K_{1A}}$ and $m_{K_{1B}} f_{K_{1B}}$ are functions
 of $\beta_{K_1}$ and quark masses only. In other words, they do not depend on
 $m_{K_{1A, 1B}}$ and hence $\theta_{K_1}$. Eq. (\ref{eq:K1decayConstMixing})
 leads to the following relation:
 \be \label{eq:K1MassDecayConst}
 m^2_{K_1 (1270)} f^2_{K_1 (1270)} +  m^2_{K_1 (1400)} f^2_{K_1 (1400)} =
  m^2_{K_{1A}} f^2_{K_{1A}}  + m^2_{K_{1B}} f^2_{K_{1B}}.
  \en
This relation, being independent of the mixing angle $\theta_{K_1}$, has been
 used \cite{Cheng10} to determine the central value of the parameter
 $\beta_{K_1}$ to be 0.3224 GeV. However, to calculate the individual decay constants,
 masses of $K_1$ mesons are needed for which the mixing angle  $\theta_{K_1}$ is required.
From Eq. (\ref{eq:K1mixing}), the masses of the $K_{1A}$ and $K_{1B}$ can be expresses as
 \be \label{eq:K1masses}
 m^2_{K_{1A}} =  m^2_{K_1 (1270)} \sin^2\theta_{K_1} + m^2_{K_1 (1400)} \cos^2\theta_{K_1},
  \nonumber\\
   m^2_{K_{1B}} =  m^2_{K_1 (1270)} \cos^2\theta_{K_1} + m^2_{K_1 (1400)} \sin^2\theta_{K_1}.
 \en

 There exists several estimations on the mixing angle in the literature
 \cite{Suzuki, Burakovsky, Cheng03}  differing in the value and sign convention.
  These often employ masses, partial decay rates of $K_1 (1270)$ and $K_1 (1400)$,
  and $\tau$ decay rates to these mesons.\footnote{The relative signs of the decay
  constants, form factors, and mixing angles of the axial vector mesons were often
  confusing in the literature. The sign of mixing angle is intimately related to the
   relative sign of the $K_{1A}$ and $K_{1B}$ states. For a detailed discussion, refer
    to \cite{Cheng07, Cheng10, Cheng08}.} Note that in the CLF quark model, the sign
    of $f_{K_{1A}}$ is negative, whereas $f_{K_{1B}}$ is positive. With this
     sign convention, from Eq. (\ref{eq:K1decayConstMixing}), the following two
      solutions \cite{Cheng10} have been obtained:
 \be \label{eq:K1mixingValue}
   \theta_{K_1}  =  \left\{  \begin{array}{c}
             +50.8^\circ ~ \textrm{solution~ I}, \\
              -44.8^\circ ~ \textrm{solution ~II}.
                    \end{array} \right.
   \en
The second solution is ruled out by the experimental data for $ B \rightarrow K_1 (1270)/K_1 (1400)
 + \gamma$ decays \cite{Cheng10}. For $\theta_{K_1}  = 50.8^\circ$, masses of  $^3P_1$ and $^1P_1$
 states come out to be,
\be \label{eq:K1ABmasses}
m_{K_{1A}} = 1.26 ~\textrm{GeV}, ~~ m_{K_{1B}} = 1.52~\textrm{ GeV},
   \en
corresponding to
\be \label{eq:K1decConstPhy}
 f_{K_1(1270)}= -170~ \textrm{MeV}, ~~  f_{K_1(1400)} = 139~ \textrm{MeV}.
 \en
Note that obtained value for $\beta_{K_1}$ lies between $\beta_{K}$ and $\beta_{K^*}$.

(x) Like $s$-wave mesons, there are mixing between singlet
and octet states of $p$-wave mesons also, equivalently, between $u\overline{u} + d \overline{d}$ and
 $s \overline{s}$ components. For axial vector states $f_1 (1285)$ and $f_1 (1420)$,
 the mixing can be written as
 \be \label{eq:f1mixing}
 f_1(1285)=f_{1q} \sin\alpha_{f_1} +   f_{1s}\cos\alpha_{f_1},
 \nonumber\\
  f_1(1420)=f_{1q} \cos\alpha_{f_1} -  f_{1s}\sin\alpha_{f_1},
 \en
 where
$f_{1q}= (u\overline{u} + d\overline{d})/\surd {2}$, and
$f_{1s}$ is pure $(s\overline{s})$ state. The mixing angle $\alpha_{f_1} $
is related to the singlet-octet mixing angle
$\theta_{f_1}$ by the  $ \alpha_{f_1} = \theta_{f_1} + 54.7^\circ$,
 where the latter mixing angle is defined by
\be \label{eq:f1mixing18}
 f_1(1285)=  f_1 \cos\theta_{f_1} + f_8 \sin\theta_{f_1} ,
 \nonumber\\
  f_1(1420)= - f_1 \sin\theta_{f_1} + f_8 \cos\theta_{f_1} .
 \en
The magnitude of the angle is given by the mass relations
\be \label{eq:f1mixingAngle}
 \tan^2 \theta_{f_1}  ={ { 4 m^2_{K_{1A}} - m^2_{a_1} - 3 m^2_{f_1 (1420)} } \over
{- 4 m^2_{K_{1A}} + m^2_{a_1} + 3 m^2_{f_1 (1285)}} },
\en
while the sign of the angle can be determined from
\be \label{eq:f1mixingAngle1}
\tan \theta_{f_1} ={ { 4 m^2_{K_{1A}} - m^2_{a_1} - 3 m^2_{f_1 (1420)}} \over
{2 \surd {2}( m^2_{a_1} - m^2_{K_{1A}}) } }.
\en
We thus obtain $\alpha_{f_1} = 94.9^\circ$, i.e., $\theta_{f_1} = 40.2^\circ$.
  Denoting the mass of the $f_{1s}$ component as $m_{f_{1s}}$, we have
 \be \label{eq:f1ssmass}
m^2_{f_{1s}} = m^2_{f_1 (1420)} \sin^2 \alpha_{f_1} + m^2_{f_1 (1285)} \cos^2 \alpha_{f_1},
\en
which yields $m_{f_{1s}}= 1.425$ GeV. Using the mixing angle and $m_{s \overline{s}}$,
the decay constant  $f_{f_{1s}}$ of the $^3 P_1$ axial vector meson with a pure
 $s \overline{s}$ quark content has been determined to be $-230\pm 9 $ MeV \cite{Yang10}.
Consequently, $\beta_{f_{1s} }$ gets fixed in the present CLF model to
be $0.3492\pm0.0064$ \cite{Cheng10}. For the purpose
of an estimation, for the remaining axial vector mesons,  we use the $\beta$ parameters
obtained in the ISGW2 model~\cite{ISGW2} up to some simple scaling, and $10 \%$ uncertainty
has been assigned to them arbitrarily to study its effects on their decay constants and
the corresponding form factors.

(xi) The $\beta$ values are kept same for other
$p$-wave mesons, scalar $(J^{PC}=0^{++})$ and axial vector
$(J^{PC}=1^{+-})$  mesons, as that of  the $(J^{PC}=1^{++})$  mesons
having the same flavor quantum numbers.
So their decay constants are calculated respectively as shown in Table II.
   The $\beta$ parameters for $p$-wave states of the charmed and bottom states
are smaller when compared to the respective $\beta_{P,V}$ values.

(xii)   Situation regarding the decay constant for the $^1 P_1$ mesons is different
from the $^3 P_1$ mesons. First of all, its decay
constant vanishes in the isospin or SU(3) limit. In fact, because of charge
conjugation invariance, the decay constant of the nonstrange neutral
meson $b^0_1 (1235)$ must be zero. In the isospin limit, the decay constant
 of the charged $b_1$ vanishes due to the fact that the $b_1$ has even
 $G$-parity and that the relevant weak axial-vector current is odd under
 $G$ transformation. Hence, $f_{b^+_1 (1235)}$ is very small in reality, arising due to the
small mass difference between $u$ and $d$ quark masses. In the present covariant
light-front quark model, if we increase the constituent $d$ quark mass by
an amount of $5\pm2$ MeV relative to the $u$ quark one, we find
 $f_{b^+_1 (1235)}= 0.6 \pm 0.2 $ MeV which is highly suppressed.
\footnote {In \cite{Laporta06}, the decay constants
of $a_1$ and $b_1$ are derived using the $K_{1A} -K_{1B}$ mixing angle
 $\theta_{K_1}$ and SU(3) symmetry to be $(f_{b_1}; f_{a_1}) = (74; 215)$ MeV
 for $\theta_{K_1}= 32^\circ$  and $(-28; 223)$ MeV for $\theta_{K_1}= 58^\circ$.
  It seems to us that the magnitude of $b_1$  decay constant derived
   in this manner is too big.}
Similar to the $f_1(1285) -f_1 (1420)$ mixing in the
 $J^{PC} = 1^{++}$  nonet, the $h_1 (1170)- h_1 (1380)$
 mixing can be described for $J^{PC} = 1^{+-}$ with the replacement
 $ f_1 (1285)\rightarrow h_1 (1170), f_1 (1420)\rightarrow h_1 (1380)$, and
 $\theta_{f_1} \rightarrow \theta_{h_1},  \alpha_{f_1} \rightarrow \alpha_{h_1}$,
 leading to $\alpha_{h_1} = 54.7^\circ$ \cite{Cheng10}, i.e., $\theta_{h_1} = 0^\circ$.
 However, like $b^0_1$, decay constants of these mesons also vanish.
  In fact,  in the SU(3) limit, $f_{^1 P_1} = 0$ should follow for strange mesons also
in this nonet. However, for the strange mesons $K_{1B}$, nonzero decay constant
would arise through SU(3) breaking, and we obtain $f_{K_{1B}}= 20.4^{+2.5}_{-1.8}$ MeV.
For charmed and bottom axial vector mesons, the results given in Table \ref{tab:f} translate
 to $f^{1/2}_{D_1}= 179^{+37}_{-34}, f^{3/2}_{D_1}=-53.6^{-14.0}_{+12.1},$
 $f^{1/2}_{D_{1s}}= 154^{+34}_{-31}, f^{3/2}_{D_{1s}}=-57.3^{-14.7}_{+12.8},$
  $f^{1/2}_{B_1}= 175^{+33}_{-30}, f^{3/2}_{B_1}=-21.4^{-5.7}_{+4.9},$
 $f^{1/2}_{B_{1s}}= 183^{+37}_{-34}, f^{3/2}_{B_{1s}}=-28.3^{-7.5}_{+6.4},$
  $f^{1/2}_{B_{1c}}= 157^{+37}_{-33}$, and $ f^{3/2}_{B_{1c}}=-47.3^{-12.4}_{+10.7}$
   (all in MeV).  The errors shown here occur due to the $10 \%$ arbitrary uncertainty
  assigned to their $\beta$ values. Note that the decay constants of $^3P_1$
  and $P_1^{3/2}$ states have opposite
signs to that of $^1P_1$ or $P_1^{1/2}$ as can be easily seen from
Eq. (\ref{eq:Phalf}).

(xiii) Similarly for scalar mesons, their decay constants also vanish
in the SU(N) limit, as has been shown above Eq.(\ref{eq:Seom}) by applying equations
of motion. However, due to SU(N) breaking, only off-diagonal scalar
mesons can have nonzero decay constants, which have been given in
Table \ref{tab:f}. In the present covariant
light-front quark model, if the constituent $d$ quark mass is increased by
an amount of $5\pm2$ MeV relative to the $u$ quark one, we find
 $|f_{a^{\pm}_0 (1450)}|= 1.1 \pm 0.4 $ MeV which is highly suppressed, whereas
 SU(3) breaking yields  $|f_{K^{*\pm}_0}|\approx 35$ MeV. Thus it
is clear that the decay constant of light scalar resonances remain largely suppressed
 relative to that of the pseudoscalar mesons owing to the small mass difference
  between the constituent quark masses, though this suppression
becomes less restrictive for heavy scalar mesons because of heavy
and light quark mass imbalance, and decay constants are of the order of hundred MeV.
Note that what is the underlying quark structure of light scalar
resonances is still controversial. While it has been widely
advocated that the light scalar nonet formed by $\sigma(600)$,
$\kappa(800)$, $f_0(980)$ and $a_0(980)$ can be identified
primarily as four-quark states, it is generally believed that the
nonet states $a_0(1450)$, $K_0^*(1430)$, $f_0(1370)$ and
$f_0(1500)/f_0(1710)$ are the conventional $q\bar q'$ states \cite{Close}.  Therefore, the prediction of
$f_{K^*_0} \approx 35 $ MeV for the scalar meson in the $s \bar u$ content (see
Table \ref{tab:f}) is most likely designated for the $K_0^*(1430)$
state. Notice that this prediction is slightly smaller than the
result of $42$ MeV obtained in \cite{Maltman} based on the
finite-energy sum rules, and far less than the estimate of
$(70\pm10)$ MeV  in \cite{Chernyak}. It is worth remarking that
even if the light scalar mesons are made from $4$ quarks, the decay
constants of the neutral scalars $\sigma(600)$, $f_0(980)$ and
$a_0^0(980)$ must vanish owing to charge conjugation invariance.

(xiv) In this work, we have only considered
the scalar nonet with masses above 1 GeV, for which the
quark content of $a_0(1450)$ and  $K_0^*(1430)$ is quite
obvious, whereas the internal structure of the isoscalars $f_0(1370)$,
$f_0(1500)$ and $f_0(1710)$ in the same nonet is controversial
and less clear. Since not all the three isosinglet scalars can be
accommodated in the $q\bar q$ nonet picture, one of them should be
primarily a scalar glueball . Among them, it has been quite
controversial as to which of these
is the dominant scalar glueball. It has been
advocated that $f_0(1710)$ is mainly $(s\overline{s})$ and $f_0(1500)$ mostly
gluonic \cite{Close1, Amsler10}. However, this scenario encounters several
insurmountable difficulties, see \cite{Kleefeld, HYCScalar06} for detailed
discussions. Based on two simple and robust results as the input
for the mass matrix, the analysis in \cite{HYCScalar06} shows that in the limit of
exact SU(3) symmetry, $f_0(1500)$ is an SU(3) isosinglet octet
state and is degenerate with $a_0(1450)$. In the absence of glueball-quarkonium mixing, $f_0(1370)$
becomes a pure SU(3) singlet  and $f_0(1710)$
the pure glueball.  When the
glueball-quarkonium mixing is turned on, there will be some
mixing between the glueball ($G$) and the SU(3)-singlet $q\bar{q}$.
The mixing matrix obtained in this model has
the form \cite{HYCScalar06}:
\begin{eqnarray}  \label{eq:wf}
 \left(\begin{matrix} { f_0(1370) \cr f_0(1500) \cr f_0(1710) \cr} \end{matrix}\right)=
\left( \begin{matrix} { 0.78 & 0.51 & -0.36 \cr
                 -0.54 & 0.84 & 0.03 \cr
                0.32 & 0.18 & 0.93 \cr}
                  \end{matrix}\right)\left(\begin{matrix}{f_{0q} \cr
 f_{0s} \cr G \cr}\end{matrix}\right),
\end{eqnarray}
where $f_{0q}= (u\overline{u} + d\overline{d})/\surd {2}$, and
$f_{0s}$ is pure $(s\overline{s})$ state,
with masses $1.474$ GeV and $1.5$ GeV, respectively.
It is evident that $f_0(1710)$ is composed primarily of the scalar
glueball, $f_0(1500)$ is close to an SU(3) octet, and $f_0(1370)$
consists of an approximate SU(3) singlet with some glueball
component ($\sim 10\%$). Note that the recent quenched and unquenched lattice
calculations all favor a scalar glueball
mass close to 1700 MeV \cite{ChenMorningstar}.

(xv)   In principle, the decay constant of the scalar strange charmed
meson $D^*_{s0}$ can be determined from the hadronic decay $B\to
\ov DD_{s0}^*$ since it proceeds only via external $W$-emission.
Indeed, a measurement of the $D\bar D_{s0}^*$ production in
$B$ decays by Belle \cite{BelleDs-1} indicates a $f_{D_{s0}^*}$ of
order 60 MeV \cite{Cheng:2003id} which is close to the calculated value
of 71 MeV (see Table \ref{tab:f}).
In our earlier work \cite{Cheng04}, we have discussed
more about $\ov DD_s^{**}$ productions in $B$ decays. The
smallness of the decay constant $f_{D_{s0}^*}$ relative to
$f_{D_s}$ can be seen from Eqs. (\ref{eq:fP}) and (\ref{eq:fS})
that
 \be
 f_{D_s(D_{s0}^*)}\propto \int dx_2\cdots[m_c x_2\pm m_s(1-x_2)].
 \en
Since the momentum fraction $x_2$ of the strange quark in the
$D_s(D_{s0}^*)$ meson is small, its effect being constructive in
$D_s$ case and destructive in $D_{s0}^*$ is sizable and explains
why $f_{D_{s0}^*}/f_{D_s}\sim 0.2$.

(xvi) For $D$ and $B$ systems, it is clear from Table~\ref{tab:f} that
$|f_{A^{3/2}}|\ll f_S <  f_{A^{1/2}}$, in accordance with the
expectation from HQS [cf. Eq. (\ref{eq:HQSf})]. Decay constants of
$p$-wave charmed and bottom mesons
have been obtained using the Bethe-Salpeter method \cite{Wang07},
 which are consistent with values obtained in Table II
 for the bottom sector and $^1 P_1$ charmed mesons,
and are slightly higher than that of other $p$-wave charmed
mesons. However, our values for $D^*_{s0}$ and $D_{s1}$ match well with the results
 $f_{D^*_{s0}}= 67.1 \pm 4.5 $ MeV and  $f_{D_{s1}}= 144.5 \pm 11.1 $ MeV
obtained in \cite{Faessler07} based on the analysis of
 $B\rightarrow D^* + D^*_{s0}/D_{s1}$ decays. For charmed $(c\overline{u})$
 meson, our estimate  $f_{D^{1/2}_1} = 179^{+37}_{-34} $ MeV is consistent with values
 $f_{D^{1/2}_1}= 196 \pm93  $ MeV and  $206 \pm120  $ MeV obtained in \cite{Jugeau05}
 from the analysis of $\overline{B} \rightarrow D^{1/2}_1 \pi $ and
 $\overline{B} \rightarrow D^{1/2}_0 \pi $ decays, respectively.

(xvii) The $\beta$ values used in the present analysis often differ from the
ones given in the earlier work ~\cite{Cheng04} to match with the
decay constants based on the latest data. For the same reason,
the strange quark mass $m_s = 0.45 $ GeV used here is different from
the values $0.37$ GeV used earlier ~\cite{Cheng04}. This choice of the strange
quark mass has to be made to obtain $f_{K^*_0} = 0.35$ MeV
based on the analysis of the $B$ decays emitting $K^*_0$
\cite{Cheng06,ChengChua10}. Otherwise, for the strange
quark mass $m_s = 0.37$ GeV, this decay constant would require $\beta_{K^*_0} > 0.60$,
which is quite high for $p$-wave mesons. Particularly, it would also spoil
the matching for decay constants of axial vector $K_1$ mesons which seem to
 require $\beta < 0.4 $. Furthermore, choosing $\beta_{K^*_0} > 0.60$, would also enhance
 $f_{D^*_{s0}}$  and  $f_{D^{1/2}_{s1}}$ unduly high if their  $\beta$'s
 are taken to be greater than or equal to that of ${K^*_0}$.
 The new choice of  $m_s = 0.45$ GeV has obviously resulted in
difference in the obtained form factors involving strange mesons
 from that given in the earlier work ~\cite{Cheng04}.

(xviii) In this work, we have investigated the variation of the form factors
and their slope parameters for $q^2$ dependence with the variation of
$\beta$ values. Wherever the experimental information is available, we have
used that to fix the errors in the beta values,
otherwise a standard $10 \%$  uncertainty in $\beta$ is assigned to
the remaining $s$-wave and $p$-wave mesons.

\section{Covariant model analysis of form factors}

In this section we first describe the form factors
for $s$-wave mesons within the framework of the covariant
light-front quark model \cite{Jaus99} and then extend it to the
$p$-wave meson case followed by numerical results and discussions in the next section.

\subsection{Form factors for $s$-wave to $s$-wave transitions}

Form factors for $P\to P,V$ transitions are defined by
  \be
\la P(P^\pp)|V_\mu|P(P^\prime)\ra
          &=& P_\mu f_+(q^2) + q_\mu f_-(q^2),
\non\\
 \la V(P^\pp,\vp^\pp)|V_\mu|P(P^\prime)\ra
          &=&\epsilon_{\mu\nu\alpha \beta}\,\vp^{\pp*\nu}P^\alpha q^\beta\, g({q^2}),
\non\\
\la V(P^\pp,\vp^\pp)|A_\mu|P(P^\prime)\ra
          &=&-i\left\{\varepsilon_\mu^{\pp*} f({q^2})
              +\vp^{*\pp}\cdot P \left[P_\mu a_+({q^2})+q_\mu a_-({q^2})\right]\right\},
 \label{eq:ffs}
 \en
where $P=P^\prime+P^\pp$, $q=P^\prime-P^\pp$ and the convention
$\epsilon_{0123}=1$ is adopted. These form factors are related to
the commonly used Bauer-Stech-Wirbel (BSW) form factors \cite{BSW}
via
 \be \label{eq:ffsdimless}
 F_1^{PP}(q^2)&=&f_+(q^2),\quad
                 F_0^{PP}(q^2)=f_+(q^2)+\frac{q^2}{q\cdot P} f_-(q^2),
 \non\\
 V^{PV}(q^2)&=&-(M^\prime+M^\pp)\, g(q^2),\quad
                 A_1^{PV}(q^2)=-\frac{f(q^2)}{M^\prime+M^\pp},
 \non\\
 A_2^{PV}(q^2)&=&(M^\prime+M^\pp)\, a_+(q^2),\quad
                 A_3^{PV}(q^2)-A_0^{PV}(q^2)=\frac{q^2}{2 M^\pp}\, a_-(q^2),
 \en
 where the latter form factors are defined by \cite{BSW}
  \be \label{eq:ffPPV}
 \la P(P^\pp)|V_\mu|P(P^\prime)\ra &=& \left(P_\mu-{M^{\prime 2}-M^{\pp 2}\over q^2}\,q_ \mu\right)
F_1^{PP}(q^2)+{M^{\prime 2}-M^{\pp 2}\over q^2}\,q_\mu\,F_0^{PP}(q^2), \non \\
 \la V(P^\pp,\vp^\pp)|V_\mu|P(P^\prime)\ra &=& -{1\over
M'+M^\pp}\,\epsilon_{\mu\nu\alpha \beta}\vp^{\pp*\nu}P^\alpha
q^\beta  V^{PV}(q^2),   \non \\
 \la V(P^\pp,\vp^\pp)|A_\mu|P(P^\prime)\ra &=& i\Big\{
(M'+M^\pp)\vp^{\pp*}_\mu A_1^{PV}(q^2)-{\vp^{\pp*}\cdot P\over
M'+M^\pp}\,P_\mu A_2^{PV}(q^2)    \non \\
&& -2M^\pp\,{\vp^{\pp*}\cdot P\over
q^2}\,q_\mu\big[A_3^{PV}(q^2)-A_0^{PV}(q^2)\big]\Big\},
 \en
with $F_1^{PP}(0)=F_0^{PP}(0)$, $A_3^{PV}(0)=A_0^{PV}(0)$, and
 \be
A_3^{PV}(q^2)=\,{M'+M^\pp\over
2M^\pp}\,A_1^{PV}(q^2)-{M'-M^\pp\over 2M^\pp}\,A_2^{PV}(q^2).
 \en
Besides the dimensionless form factors,
this parametrization has the advantage that the $q^2$ dependence
of the form factors is governed by the resonances of the same
spin, for instance, the momentum dependence of $F_0(q^2)$ is
determined by scalar resonances.

To obtain the $P\to M$ transition form factors with $M$ being a
ground-state $s$-wave meson (or a low-lying $p$-wave meson), we
follow \cite{Jaus99, Cheng04} to obtain $P\to P,V$ form factors before
considering the $p$-wave meson case.
For the case of $M=P$, it is straightforward to obtain the form
factors $f_\pm(q^2)$ for $q^2=-q_\bot^2\leq0$. We will return to
 the issue of the momentum dependence of form factors in the last sub-section.  At $q^2=0$,
the form factor $f_+(0)$ is simply given by
  \be \label{eq:fplus}
 f_+(q^2)&=&\frac{N_c}{16\pi^3}\int dx_2 d^2p^\prime_\bot
            \frac{h^\prime_P h^\pp_P}{x_2 \hat N_1^\prime \hat N^\pp_1}
            \Bigg[x_1 (M_0^{\prime2}+M_0^{\pp2})+x_2 q^2
 \non\\
         &&\qquad-x_2(m_1^\prime-m_1^\pp)^2 -x_1(m_1^\prime-m_2)^2-x_1(m_1^\pp-m_2)^2\Bigg].
 \non\\
 \en
Similarly, we have
\be
 f_-(q^2)&=&\frac{N_c}{16\pi^3}\int dx_2 d^2p^\prime_\bot
            \frac{2h^\prime_P h^\pp_P}{x_2 \hat N_1^\prime \hat N^\pp_1}
            \Bigg\{- x_1 x_2 M^{\prime2}-p_\bot^{\prime2}-m_1^\prime m_2
                  +(m_1^\pp-m_2)(x_2 m_1^\prime+x_1 m_2)
\non\\
         &&\qquad +2\frac{q\cdot P}{q^2}\left(p^{\prime2}_\bot+2\frac{(p^\prime_\bot\cdot q_\bot)^2}{q^2}\right)
                  +2\frac{(p^\prime_\bot\cdot q_\bot)^2}{q^2}
                  -\frac{p^\prime_\bot\cdot q_\bot}{q^2}
                  \Big[M^{\pp2}-x_2(q^2+q\cdot P)
\non\\
         &&\qquad -(x_2-x_1) M^{\prime2}+2 x_1 M_0^{\prime
                  2}-2(m_1^\prime-m_2)(m_1^\prime+m_1^\pp)\Big]
           \Bigg\}.
 \label{eq:fpm}
 \en

We next turn to the $P\to V$ transition form factors, which are given by

\be \label{eq:PtoV}
 g(q^2)&=&-\frac{N_c}{16\pi^3}\int dx_2 d^2 p^\prime_\bot
           \frac{2 h^\prime_P h^\pp_V}{x_2 \hat N^\prime_1 \hat N^\pp_1}
           \Bigg\{x_2 m_1^\prime+x_1 m_2+(m_1^\prime-m_1^\pp)
           \frac{p^\prime_\bot\cdot q_\bot}{q^2}
           +\frac{2}{w^\pp_V}\left[p^{\prime2}_\bot+\frac{(p^\prime_\bot\cdot q_\bot)^2}{q^2}\right]
           \Bigg\},
 \non\\
  f(q^2)&=&\frac{N_c}{16\pi^3}\int dx_2 d^2 p^\prime_\bot
            \frac{ h^\prime_P h^\pp_V}{x_2 \hat N^\prime_1 \hat N^\pp_1}
            \Bigg\{2
            x_1(m_2-m_1^\prime)(M^{\prime2}_0+M^{\pp2}_0)-4 x_1
            m_1^\pp M^{\prime2}_0+2x_2 m_1^\prime q\cdot P
 \non\\
         &&+2 m_2 q^2-2 x_1 m_2
           (M^{\prime2}+M^{\pp2})+2(m_1^\prime-m_2)(m_1^\prime+m_1^\pp)^2
           +8(m_1^\prime-m_2)\left[p^{\prime2}_\bot+\frac{(p^\prime_\bot\cdot q_\bot)^2}{q^2}\right]
 \non\\
         &&
           +2(m_1^\prime+m_1^\pp)(q^2+q\cdot
           P)\frac{p^\prime_\bot\cdot q_\bot}{q^2}
           -4\frac{q^2 p^{\prime2}_\bot+(p^\prime_\bot\cdot q_\bot)^2}{q^2 w^\pp_V}
            \Bigg[2 x_1 (M^{\prime2}+M^{\prime2}_0)-q^2-q\cdot P
 \non\\
         &&-2(q^2+q\cdot P)\frac{p^\prime_\bot\cdot
            q_\bot}{q^2}-2(m_1^\prime-m_1^\pp)(m_1^\prime-m_2)
            \Bigg]\Bigg\},
 \non\\
a_+(q^2)&=&\frac{N_c}{16\pi^3}\int dx_2 d^2 p^\prime_\bot
            \frac{2 h^\prime_P h^\pp_V}{x_2 \hat N^\prime_1 \hat N^\pp_1}
            \Bigg\{(x_1-x_2)(x_2 m_1^\prime+x_1 m_2)-[2x_1
            m_2+m_1^\pp+(x_2-x_1)
            m_1^\prime]\frac{p^\prime_\bot\cdot q_\bot}{q^2}
 \non\\
         &&-2\frac{x_2 q^2+p_\bot^\prime\cdot q_\bot}{x_2 q^2
            w^\pp_V}\Big[p^\prime_\bot\cdot p^\pp_\bot+(x_1 m_2+x_2 m_1^\prime)(x_1 m_2-x_2 m_1^\pp)\Big]
            \Bigg\},
 \non\\
 a_-(q^2)&=&\frac{N_c}{16\pi^3}\int dx_2 d^2 p^\prime_\bot
            \frac{ h^\prime_P h^\pp_V}{x_2 \hat N^\prime_1 \hat N^\pp_1}
            \Bigg\{2(2x_1-3)(x_2 m_1^\prime+x_1
            m_2)-8(m_1^\prime-m_2)\left[\frac{p^{\prime2}_\bot}{q^2}+2\frac{(p^\prime_\bot\cdot
            q_\bot)^2}{q^4}\right]
 \non\\
         &&-[(14-12 x_1) m_1^\prime-2 m_1^\pp-(8-12 x_1) m_2]
           \frac{p^\prime_\bot\cdot q_\bot}{q^2}
 \non\\
         &&+\frac{4}{w^\pp_V}\Bigg([M^{\prime2}+M^{\pp2}-q^2+2(m_1^\prime-m_2)(m_1^\pp+m_2)]
                                   (A^{(2)}_3+A^{(2)}_4-A^{(1)}_2)
\non\\
         &&                         +Z_2(3 A^{(1)}_2-2A^{(2)}_4-1)
                                    +\frac{1}{2}[x_1(q^2+q\cdot P)
                                                 -2 M^{\prime2}-2 p^\prime_\bot\cdot q_\bot
\non\\
         &&                                      -2 m_1^\prime(m_1^\pp+m_2)-2
                                                 m_2(m_1^\prime-m_2)
                                                 ](A^{(1)}_1+A^{(1)}_2-1)
\non\\
        &&                          +q\cdot P\Bigg[\frac{p^{\prime2}_\bot}{q^2}
                                                  +\frac{(p^\prime_\bot\cdot q_\bot)^2}{q^4}\Bigg]
                                                  (4
                                                  A^{(1)}_2-3)
                             \Bigg)
            \Bigg\},
 \en

 where various quantities appearing in these formulas have been described in the previous section.

\subsection{Form factors for $s$-wave to $p$-wave transitions}

The general expressions for $P$ to low-lying $p$-wave meson
transitions are given by \cite{ISGW}
 \be \label{ISGWform}
 \la S(P^\pp)|A_\mu|P(P^\prime)\ra &=& i\Big[ u_+(q^2)P_\mu+u_-(q^2)q_\mu
 \Big], \non \\
 \la A^{1/ 2}(P^\pp,\vp^\pp)|V_\mu|P(P^\prime)\ra
           &=& i\left\{\ell_{1/2}(q^2)\vp_\mu^{\pp*}+\vp^{\pp*}\cdot
                  P[P_\mu c_+^{1/2}(q^2)+q_\mu c_-^{1/2}(q^2)]\right\},
 \non \\
 \la A^{1/2}(P^\pp,\vp^\pp)|A_\mu|P(P^\prime)\ra
           &=& -q_{1/2}(q^2)\epsilon_{\mu\nu\alpha\beta}\vp^{\pp*\nu}P^\alpha
q^\beta,
 \non \\
 \la A^{3/ 2}(P^\pp,\vp^\pp)|V_\mu|P(P^\prime)\ra
           &=& i\left\{\ell_{3/2}(q^2)\vp_\mu^{\pp*}+\vp^{\pp*}\cdot
               P[P_\mu c_+^{3/2}(q^2)+q_\mu c_-^{3/2}(q^2)]\right\},
 \non \\
  \la A^{3/2}(P^\pp,\vp^\pp)|A_\mu|P(P^\prime)\ra
           &=& -q_{3/2}(q^2)\epsilon_{\mu\nu\alpha\beta}\vp^{\pp*\nu}P^\alpha
q^\beta.
 \label{eq:ffp}
  \en
The form factors $\ell_{1/2(3/2)},c_+^{1/2(3/2)},c_-^{1/2(3/2)}$
and $q_{1/2(3/2)}$ are defined for the transitions to the heavy
$P^{1/2}_1$ ($P^{3/2}_1$) state. For transitions to light
axial-vector mesons, it is more appropriate to employ the $L-S$
coupled states $^1P_1$ and $^3P_1$ denoted by the particles
$^1\!A$ and $^3\!A$ in our notation. The relation between
$P^{1/2}_1,P^{3/2}_1$ and $^1P_1,\,^3P_1$ states is given by Eq.
(\ref{eq:Phalf}). The corresponding form factors
$\ell_{^1\!A(^3\!A)},c_+^{^1\!A(^3\!A)},c_-^{^1\!A(^3\!A)}$ and
$q_{^1\!A(^3\!A)}$ for $P\to \,^1\!A$ ($^3\!A$) transitions can be
defined in an analogous way.\footnote{The form factors
$\ell_{^1\!A(^3\!A)},c_+^{^1\!A(^3\!A)},c_-^{^1\!A(^3\!A)}$ and
$q_{^1\!A(^3\!A)}$ are dubbed as $\ell(v),c_+(s_+),c_-(s_-)$ and
$q(r)$, respectively, in the ISGW model \cite{ISGW}.}

Note that only the form factors $u_+(q^2),u_-(q^2)$ and $k(q^2)$
in the above parametrization are dimensionless. It is thus
convenient to define dimensionless form factors by\footnote{The
definition here for dimensionless $P\to A$ transition form factors
differs than Eq. (3.17) of \cite{Cheng:2003id} where the
coefficients $(m_P\pm m_A)$ are replaced by $(m_P\mp m_A)$. It
has been made clear in the earlier work \cite{Cheng04}
that this definition will lead to HQS
relations for $B\to D_0^*,D_1$ transitions   similar to that for $B\to D,D^*$ ones.}
 \be \label{eq:ffpdimless}
 \la S(P^\pp)|A_\mu|P(P^\prime)\ra &=&
 -i\left[\left(P_\mu-{M^{\prime 2}-M^{\pp 2}\over q^2}\,q_
 \mu\right) F_1^{PS}(q^2)+{M^{\prime 2}-M^{\pp 2}\over q^2}
 \,q_\mu\,F_0^{PS}(q^2)\right], \non \\
 \la A(P^\pp,\vp^\pp)|V_\mu|P(P^\prime)\ra &=&
-i\Bigg\{(m_P-m_A) \vp^*_\mu V_1^{PA}(q^2)  - {\vp^*\cdot
P'\over m_P-m_A}P_\mu V_2^{PA}(q^2) \non \\
&-& 2m_A {\vp^*\cdot P'\over
q^2}q_\mu\left[V_3^{PA}(q^2)-V_0^{PA}(q^2)\right]\Bigg\},
\non \\
   \la A(P^\pp,\vp^\pp)|A_\mu|P(P^\prime)\ra &=& -{1\over
  m_P-m_A}\,\epsilon_{\mu\nu\rho\sigma}\vp^{*\nu}P^\rho
  q^{\sigma}A^{PA}(q^2),
 \en
with
 \be V_3^{PA}(q^2)=\,{m_P-m_A\over 2m_A}\,V_1^{PA}(q^2)-{m_P+m_A\over
2m_A}\,V_2^{PA}(q^2),
 \en
and $V_3^{PA}(0)=V_0^{PA}(0)$. They are related to the form
factors in (\ref{eq:ffPPV}) via
 \be
 F^{PS}_1(q^2)&=&-u_+(q^2),\quad
                 F^{PS}_0(q^2)=-u_+(q^2)-\frac{q^2}{q\cdot P} u_-(q^2),
 \non\\
 A^{PA}(q^2)&=&-(M^\prime-M^\pp)\, q(q^2),\quad
                 V^{PA}_1(q^2)=-\frac{\ell(q^2)}{M^\prime-M^\pp},
 \non\\
 V^{PA}_2(q^2)&=&(M^\prime-M^\pp)\, c_+(q^2),\quad
                 V^{PA}_3(q^2)-V^{PA}_0(q^2)=\frac{q^2}{2 M^\pp}\, c_-(q^2).
 \en
In above equations, the axial-vector meson $A$ stands for
$A^{1/2}$ or $A^{3/2}$.

The $P\to S\,(A)$ transition form factors can be easily
obtained by some suitable modifications on $P\to P\,(V)$ ones.
The $P\to S$ transition form factors are related to $f_\pm$ by
  \be 
  u_\pm=-f_\pm(m_1^\pp\to -m_1^\pp,h^\pp_P\to h^\pp_S).
 \en
Thus the following form of these form factors can be obtained
from that of $P\to P$ ones by the replacements given above,
 \be
 u_+(q^2)&=&\frac{N_c}{16\pi^3}\int dx_2 d^2p^\prime_\bot
            \frac{h^\prime_P h^\pp_S}{x_2 \hat N_1^\prime \hat N^\pp_1}
            \Big[-x_1 (M_0^{\prime2}+M_0^{\pp2})-x_2 q^2
 \non\\
         &&\qquad+x_2(m_1^\prime+m_1^\pp)^2 +x_1(m_1^\prime-m_2)^2+x_1(m_1^\pp+m_2)^2\Big],
 \non\\
 u_-(q^2)&=&\frac{N_c}{16\pi^3}\int dx_2 d^2p^\prime_\bot
            \frac{2h^\prime_P h^\pp_S}{x_2 \hat N_1^\prime \hat N^\pp_1}
            \Bigg\{ x_1 x_2 M^{\prime2}+p_\bot^{\prime2}+m_1^\prime m_2
                  +(m_1^\pp+m_2)(x_2 m_1^\prime+x_1 m_2)
\non\\
         &&\qquad -2\frac{q\cdot P}{q^2}\left(p^{\prime2}_\bot+2\frac{(p^\prime_\bot\cdot q_\bot)^2}{q^2}\right)
                  -2\frac{(p^\prime_\bot\cdot q_\bot)^2}{q^2}
                  +\frac{p^\prime_\bot\cdot q_\bot}{q^2}
                  \Big[M^{\pp2}-x_2(q^2+q\cdot P)
\non\\
         &&\qquad -(x_2-x_1) M^{\prime2}+2 x_1 M_0^{\prime
                  2}-2(m_1^\prime-m_2)(m_1^\prime-m_1^\pp)\Big]
           \Bigg\}.
 \label{eq:upm}
 \en

Similarly, the analytic expressions for $P\to A$ transition form factors
can be obtained from that of $P\to V$ ones by the following
replacements:

 \be \label{eq:PtoA}
 \ell^{^3\!A,^1\!A}(q^2)&=&f(q^2) \,\,\,{\rm with}\,\,\,
                         (m_1^\pp\to -m_1^\pp,\,h^\pp_V\to h^\pp_{^3\!A,^1\!A},\,w^\pp_V\to w^\pp_{^3\!A,^1\!A}),
 \non\\
 q^{^3\!A,^1\!A}(q^2)&=&g(q^2) \,\,\,{\rm with}\,\,\,
                         (m_1^\pp\to -m_1^\pp,\,h^\pp_V\to h^\pp_{^3\!A,^1\!A},\,w^\pp_V\to  w^\pp_{^3\!A,^1\!A}),
 \non\\
 c_+^{^3\!A,^1\!A}(q^2)&=&a_+(q^2) \,\,\,{\rm with}\,\,\,
                          (m_1^\pp\to -m_1^\pp,\,h^\pp_V\to h^\pp_{^3\!A,^1\!A},\,w^\pp_V\to  w^\pp_{^3\!A,^1\!A}),
 \non\\
 c_-^{^3\!A,^1\!A}(q^2)&=&a_-(q^2) \,\,\,{\rm with}\,\,\,
                           (m_1^\pp\to -m_1^\pp,\,h^\pp_V\to h^\pp_{^3\!A,^1\!A},\,w^\pp_V\to  w^\pp_{^3\!A,^1\!A}).
 \en
It should be cautious that the replacement of $m_1^\pp\to
-m_1^\pp$ should not be applied to $m_1^\pp$ in $w^\pp$ and
$h^\pp$. These form factors can be expressed in the $P^{3/2}_1$
and $P^{1/2}_1$ basis by using Eq.~(\ref{eq:Phalf}). For further details,
 the reader is referred to the earlier work \cite{Cheng04}.

\subsection{Form-factor momentum dependence and numerical results}

Because of the condition $q^+=0$ we have imposed during the course
of calculation, form factors are known only for spacelike momentum
transfer $q^2=-q^2_\bot\leq 0$, whereas only the timelike form
factors are relevant for the physical decay processes. It has been
proposed in \cite{Jaus96} to recast the form factors as explicit
functions of $q^2$ in the spacelike region and then analytically
continue them to the timelike region. Another approach is to
construct a double spectral representation for form factors at
$q^2<0$ and then analytically continue it to $q^2>0$
region~\cite{Melikhov96}. It has been shown recently that, within
a specific model, form factors obtained directly from the timelike
region (with $q^+>0$) are identical to the ones obtained by the
analytic continuation from the spacelike region~\cite{BCJ03}.

In principle, form factors at $q^2>0$ can be evaluated directly in
the frame where the momentum transfer is purely longitudinal,
i.e., $q_\bot=0$, so that $q^2=q^+q^-$ covers the entire range of
momentum transfer \cite{Cheng97}. The price one has to pay is
that, besides the conventional valence-quark contribution, one
must also consider the non-valence configuration (or the so-called
$Z$-graph) arising from quark-pair creation from the vacuum.
However, a reliable way of estimating the $Z$-graph contribution
is still lacking unless one works in a specific model, for
example, the one advocated in \cite{BCJ03}. Fortunately, this
additional non-valence contribution vanishes in the frame where
the momentum transfer is purely transverse i.e., $q^+=0$.

To proceed we find that, except for the form factor $V_2$ to be
discussed below, the momentum dependence of form factors in the
spacelike region can be well parameterized and reproduced in the following
three-parameter form:
 \be
 \label{eq:FFpara}
 F(q^2)= { F(0)\over {1-a(q^2/m_{B(D)}^2)+b(q^2/m_{B(D)}^2)^2}} ,
 \en
 for $P \to M$ transitions, where $F$ stands for the relevant form factors
 appearing in these transitions.

The parameters $a$, $b$ and $F(0)$ are first determined in the
spacelike region. We then employ this parametrization to determine
the physical form factors at $q^2\geq 0$. In practice, the
parameters $a,b$ and $F(0)$ are obtained by performing a
5-parameter fit to the form factors in the range $-20\,{\rm
GeV}^2\leq q^2\leq 0$ for $B$ decays and $-10\,{\rm GeV}^2\leq
q^2\leq 0$ for $D$ decays. All $P \rightarrow M$ form factors are
calculated at five $q^2$ values given below:

a) for the charm sector:   $q^2  =  -0.01, -0.1, -1.0, -5.0, -10.0 ~ \textrm{GeV}^2 $,

b) for the bottom sector:  $q^2 =  -0.01, -0.1, -5.0, -10.0, -20.0 ~ \textrm{GeV}^2 $.

These parameters are generally
insensitive to the $q^2$ range to be fitted except for the form
factor $V_2(q^2)$ in $B(D)\to\, ^1P_1,P_1^{3/2}$ transitions.
The obtained $a$ and $b$ coefficients are in most
cases not far from unity as expected.

We have also analyzed the sensitivity of the form factors $F(0)$, and the
slope parameters ($a$ and $b$) to the uncertainties
of $\beta$ values.  The form factors at $q^2 = 0 $ are generally found to be
less sensitive to
the variation in $\beta$ values, whereas the corresponding parameters $a$ and $b$
are rather sensitive to the chosen range for $\beta$.  Numerical results
and discussion of these form factors and slope parameters ($a$ and $b$)
 are presented in detail in the following section.

\section{Numerical results and discussion}

Equipped with the explicit expressions of the form
factors $f_+(q^2),f_-(q^2)$ [ Eqs. (\ref{eq:fplus}) and (\ref{eq:fpm})] for $P\to P$
transitions, $g(q^2),f(q^2),a_+(q^2),a_-(q^2)$ [Eq.
(\ref{eq:PtoV})] for $P\to V$ transitions,
 $u_+(q^2),u_-(q^2)$ [Eq. (\ref{eq:upm})] for $P\to S$
transitions, and
$\ell(q^2),q(q^2),c_+(q^2),c_-(q^2)$ [Eq. (\ref{eq:PtoA})] for
$P\to A$ transition, we now proceed to perform their
numerical studies. In the earlier work, results for the
 form factors for $D (c \bar u)$ ,
$B(b \bar u) \to $ isovector ($\pi$ like) and isospinor
($K$ and $D$ like) transition were calculated. In this work,
we include isoscalar initial and final state mesons as well. Besides giving the updated
 results of these transitions due to the change in the strange
 quark mass, the variation of the $\beta$ values and performing fit for five $q^2$
  values, the $D_s, B_s \to P, V, S$, and $A$ transition form factors are the
main new results in this work.

In Tables~\ref{tab:P2P}$-$\ref{tab:B2D},
we present calculated form factors and their $q^2$ dependence, along with their
allowed range due to uncertainties in $\beta$ values of the initial
and final mesons,  for the $P (0^-) \to P(0^-), V(1^-), S(0+), A(1^+ ~ ^3P_1)$,
and $A(1^+ ~ ^1P_1)$ transitions of the charmed $D, D_s$ and bottom $B, B_s$ mesons.
In calculations, we have taken the meson masses from the Particle Data Group
 \cite{PDG10}. Taking the natural flavor basis for isoscalar states of all the mesons (M), i.e.,
 $M_{q} = (u\overline{u} + d\overline{d})/\surd {2}$ and $M_s = (s\overline{s})$,
we use the following masses (in GeV): $m_{\eta_q} = 0.741$ and $m_{\eta_s} = 0.802$
for pseudoscalar mesons taken from an analysis given in \cite{ChengChua10},
$m_{f_{1q}} = 1.283$ and $m_{f_{1s}} = 1.425$ for the axial-vector ($1^{++}$) nonet,
$m_{h_{1q}} = 1.242$ and $m_{h_{1s}} = 1.314$ for the other axial-vector ($1^{+-}$) case, and
$m_{f_{0q}} = 1.474$ and $m_{f_{0s}} = 1.5$
for the scalar ($0^{++}$) mesons,
based on the respective mixing schemes described in Sec. III. Form factors for
 transitions to the physical isosinglet diagonal states
can be obtained from the Tables by including suitable Clebsch-Gordan coefficients.
For instance,
\be \label{eq:FFf1CG}
 A^{B_s f_1 (1420)} =  - \sin \alpha_{f_1} A^{B_s f_{1s}},~~
 \nonumber
 A^{B_s f_1 (1285)} =  \cos\alpha_{f_1}   A^{B_s f_{1s}},
 \\
  A^{B f_1 (1420)} = {1 \over {\sqrt{2}}}  \cos \alpha_{f_1}  A^{B f_{1q}},~~
 A^{B f_1 (1285)} =  {1 \over {\sqrt{2}}} \sin \alpha_{f_1}   A^{B f_{1q}},
\en
where $\alpha_{f_1}$ has already been defined in Sec. III. The factor $\sqrt{2}$ appears
for the $B\to f_1$ form factors, since either  $u \overline{u}$ or
 $d \overline{d}$ component of $f_{1q}$
can be transited from $B$ meson via the appropriate weak current. Similarly,
only the $s \overline{s}$ components of these mesons can be transited from the $B_s$
meson. So the size of these corresponding form factors for
physical isoscalar diagonal states gets reduced by the Clebsch-Gordan coefficients. Similar procedure
can be adopted for transitions to the isosinglet diagonal states in other multiplets.

In these tables, two sets of uncertainties in the form factors,
commonly denotes as $F(0)$, and their slope parameters ($a$ and $b$) are
given. The first and second sets of uncertainties
shown in their values arise from the allowed uncertainties in the $\beta $ parameter of
the initial and final state meson,  respectively. For the sake of clarity, it is mentioned
here that the uncertainty shown as superscript (subscript) is due to the
 increase (decrease) in $\beta $ of the corresponding meson.
The obtained $a$ and $b$ coefficients are in most
cases not far from unity as expected. These parameters are generally
insensitive to the $q^2$ range to be fitted, except for the form factor
$V_2(q^2)$ in $B(D)\to\, ^1P_1,P_1^{3/2}$ transitions. For these transitions, the
corresponding parameters $a$ and $b$ are rather sensitive to the
chosen range for $q^2$, and quite larger than unity. This sensitivity is attributed to the fact
that the form factor $V_2(q^2)$ approaches to zero at very large
$-|q^2|$ where the three-parameter parametrization
(\ref{eq:FFpara}) becomes questionable. To overcome this
difficulty, we follow \cite{Cheng04} to fit this form factor to the following form:
 \be \label{eq:FFpara1}
 F(q^2)=\,{F(0)\over (1-q^2/m_{B(D)}^2)[1-a(q^2/m_{B(D)}^2)+b(q^2/m_{B(D)}^2)^2]},
 \en
and achieve a substantial improvement. For example, we have
$a=2.18$ and $b=6.08$ when $V_2^{BK_{^1P_1}}$ is fitted to Eq.
(\ref{eq:FFpara}) and they become $a=1.74$ and $b=2.17$ (see Table
\ref{tab:P2An}) when the fit formula Eq. (\ref{eq:FFpara1}) is
employed. It may be noted that we have considered parent meson constituted of
heavy quark and light antiquark, since certain decay constants and form factors
may change sign.

We make the following observations:

\subsection{ $P(0^-) \to P (0^-)$  Form Factors}

\begin{itemize}

\item From Table~\ref{tab:P2P}, we notice that heavy-to-light form factors
 for the bottom mesons are smaller (around 0.3) than all
the charmed meson form factors and the heavy-to-heavy bottom meson form
factors, $F^{BD(B_sD_s)}_{0,1}$, which are around  0.7 or 0.8.

\item We notice that the values of form factors at $q^2= 0$ for $B_s$ transitions
are similar to the corresponding ones in $B$ transitions. Therefore, flavor of
the spectator quark does not seem to play a special role in affecting them.
Particularly, we note the following for both $F_0(0)$ and $F_1(0)$:
$F^{B_s D_s} = F^{B D}$, $F^{B_s K}\approx F^{B \pi}$, and
$F^{B_s \eta_s}\approx F^{B \eta_q}$,  where
$\eta_{q}= (u\overline{u} + d\overline{d})/\surd {2}$, and
$\eta_{s}$ is pure $(s\overline{s})$ state. For the charm sector also, one may
 notice $F^{D_s K} = F^{D \pi} (\approx F^{BD})$ and $F^{D_s \eta_s}\approx F^{D K}$.
However, the slope parameters, $a$ and $b$, differ for these cases.

\item Since the decay constants of pseudoscalar mesons are quite accurately
determined, the errors on the $\beta$ parameters are rather small.
Correspondingly, the errors in the calculated form factors at $q^2 = 0$ are
 also very small. The same is true for the slope parameters except for a
few cases, particularly for $b$, which may show large variation sometimes.

\item Form factors $(F^{PP}_1(0)$ and $F^{PP}_0(0))$ usually tend to decrease (increase)
 with increasing (decreasing) $\beta$ for initial meson, whereas they tend to increase (decrease)
with increasing (decreasing) $\beta$ for final meson. Only for $B_s$, the form factors
show increasing (decreasing) trend for the initial as well as the final meson.

\item Usually all the slope parameters are found to be positive.
For the bottom sector, the slope parameters are larger than that for the charm sector.
Particularly, the parameter $b$ is much small ($< 0.1$, if not zero) for $F^{PP}_0(0)$, except for
$F^{B_s K}_0(0)$ and $F^{B_s \eta_s}_0(0)$ for which $b \approx 0.35$. For
$F^{B_s K}_1(0)$ and $F^{B_s \eta_s}_1(0)$, the parameter $b$ is around $2-3$
times larger than that for other cases.

\item	Slope parameters obtained using Eq. (\ref{eq:FFpara}) generally
tend to increase (decrease) with decrease (increase) in $\beta$ for
 each of the initial and final mesons.

\item According to the three-parameter parametrization Eq. (\ref{eq:FFpara}), the dipole behavior
corresponds to $b=(a/2)^2$, while $b=0$ and $a\neq0$ induces
a monopole dependence. An inspection of Table~\ref{tab:P2P}
indicates that form factors $F^{PP}_0$ generally show a monopole behavior, and $F^{PP}_1$
have a dipole behavior particularly for charmed meson transitions.

\end{itemize}

\begin{table}[b]
\caption{Form factors of
$P(0^-)\rightarrow P(0^-)$ transitions
obtained in the covariant light-front model are fitted to the
3-parameter form Eq. (\ref{eq:FFpara}). All the form factors are dimensionless. }
 \label{tab:P2P}
\begin{ruledtabular}
\begin{tabular}{|cccc|cccc|}
$F$
    & $F(0)$
    & $a$
    & $b$
    & $F$
    & $F(0)$
    & $a$
    & $b$
    \\
\hline
$F^{D  \pi}_1$
    & $0.66^{-0.01+0.00}_{+0.01-0.00}$
    & $1.19^{-0.01-0.00}_{+0.01+0.00}$
    & $0.35^{-0.03-0.00}_{+0.03+0.00}$
    & $F^{D  \pi}_0$
    & $0.66^{-0.01+0.00}_{+0.01-0.00}$
    & $0.51^{-0.00-0.00}_{+0.00+0.00}$
    & $0.00^{-0.01-0.00}_{+0.01+0.00}$
    \\
$F^{D {\eta}_q}_1$
    & $0.71^{-0.00+0.01}_{+0.00-0.01}$
    & $1.13^{-0.01-0.02}_{+0.01+0.02}$
    & $0.27^{-0.02-0.02}_{+0.02+0.02}$
    & $F^{D {\eta}_q}_0$
    & $0.71^{-0.01+0.01}_{+0.00-0.01}$
    & $0.43^{+0.01-0.03}_{-0.01+0.03}$
    & $-0.01^{-0.00+0.00}_{+0.01-0.00}$
    \\
$F^{DK}_1$
    & $0.79^{-0.01+0.00}_{+0.01-0.00}$
    & $1.05^{-0.01-0.00}_{+0.01+0.00}$
    & $0.25^{-0.02-0.00}_{+0.02+0.00}$
    & $F^{DK}_0$
    & $0.79^{-0.01+0.00}_{+0.01-0.00}$
    & $0.47^{+0.00-0.01}_{-0.00+0.01}$
    & $-0.00^{-0.00-0.00}_{+0.00+0.00}$
    \\
\hline
$F^{D_s K}_1$
    & $0.66^{-0.00+0.00}_{+0.00-0.00}$
    & $1.11^{-0.00-0.00}_{+0.00+0.00}$
    & $0.48^{-0.03-0.01}_{+0.03+0.01}$
    & $F^{D_s K}_0$
    & $0.66^{-0.00+0.00}_{+0.00-0.00}$
    & $0.56^{-0.00-0.01}_{+0.00+0.01}$
    & $0.04^{-0.01-0.00}_{+0.01+0.00}$
    \\
$F^{D_s {\eta}_s}_1$
    & $0.76^{-0.00+0.02}_{+0.00-0.03}$
    & $1.02^{-0.01-0.01}_{+0.00+0.01}$
    & $0.40^{-0.02-0.05}_{+0.02+0.05}$
    & $F^{D_s {\eta}_s}_0$
    & $0.76^{-0.01+0.02}_{+0.00-0.03}$
    & $0.60^{-0.00-0.06}_{-0.00+0.05}$
    & $0.04^{-0.00-0.01}_{+0.00+0.01}$
    \\
\hline
$F^{B\pi}_1$
    & $0.25^{-0.00+0.00}_{+0.00-0.00}$
    & $1.70^{-0.03-0.00}_{+0.03+0.00}$
    & $0.90^{-0.05-0.00}_{+0.06+0.00}$
    & $F^{B\pi}_0$
    & $0.25^{-0.00+0.00}_{+0.00-0.00}$
    & $0.82^{-0.02-0.00}_{+0.02+0.00}$
    & $0.09^{-0.01-0.00}_{+0.02+0.00}$
    \\
$F^{B {\eta}_q}_1$
    & $0.29^{-0.00+0.01}_{+0.00-0.01}$
    & $1.63^{-0.02-0.02}_{+0.02+0.02}$
    & $0.74^{-0.04-0.04}_{+0.04+0.04}$
    & $F^{B {\eta}_q}_0$
    & $0.29^{-0.00+0.01}_{+0.00-0.01}$
    & $0.75^{-0.01-0.03}_{+0.01+0.03}$
    & $0.04^{-0.01-0.01}_{+0.01+0.01}$
    \\
$F^{BK}_1$
    & $0.34^{-0.00+0.00}_{+0.00-0.00}$
    & $1.60^{-0.02-0.00}_{+0.02+0.00}$
    & $0.73^{-0.04-0.01}_{+0.04+0.01}$
    & $F^{BK}_0$
    & $0.34^{-0.00+0.00}_{+0.00-0.00}$
    & $0.78^{-0.02-0.01}_{+0.02+0.01}$
    & $0.05^{-0.01-0.00}_{+0.01+0.00}$
    \\
$F^{BD}_1$
    & $0.67^{-0.00+0.01}_{+0.00-0.01}$
    & $1.22^{-0.01-0.02}_{+0.01+0.01}$
    & $0.36^{-0.01-0.01}_{+0.01+0.02}$
    & $F^{BD}_0$
    & $0.67^{-0.00+0.01}_{+0.00-0.01}$
    & $0.63^{-0.00-0.02}_{+0.00+0.02}$
    & $-0.01^{-0.01-0.00}_{+0.01+0.00}$
    \\
\hline
$F^{B_s K}_1$
    & $0.23^{+0.00+0.00}_{-0.00-0.00}$
    & $1.88^{-0.04-0.01}_{+0.04+0.01}$
    & $1.58^{-0.12-0.03}_{+0.14+0.03}$
    & $F^{B_s K}_0$
    & $0.23^{+0.00+0.00}_{-0.00-0.00}$
    & $1.05^{-0.03-0.01}_{+0.04+0.01}$
    & $0.35^{-0.05-0.02}_{+0.04+0.00}$
    \\
$F^{B_s {\eta}_s}_1$
    & $0.28^{+0.00+0.02}_{-0.00-0.02}$
    & $1.82^{-0.04-0.05}_{+0.04+0.05}$
    & $1.45^{-0.11-0.16}_{+0.13+0.18}$
    & $F^{B_s {\eta}_s}_0$
    & $0.28^{+0.00+0.02}_{-0.00-0.02}$
    & $1.07^{-0.03-0.06}_{+0.03+0.07}$
    & $0.32^{-0.04-0.06}_{+0.05+0.07}$
    \\
$F^{B_s D_s}_1$
    & $0.67^{+0.00+0.01}_{-0.01-0.01}$
    & $1.28^{-0.02-0.02}_{+0.02+0.02}$
    & $0.52^{-0.03-0.02}_{+0.03+0.02}$
    & $F^{B_s D_s}_0$
    & $0.67^{+0.00+0.01}_{-0.01-0.01}$
    & $0.69^{-0.01-0.02}_{+0.00+0.02}$
    & $0.07^{-0.01-0.01}_{+0.01+0.01}$
     \\
\end{tabular}
\end{ruledtabular}
\end{table}

\subsection{$P(0^-) \to V (1^-)$  Form Factors}

\begin{itemize}

\item Like $P(0^-) \to P (0^-)$  form factors, we note
 the following behavior from Table~\ref{tab:P2V}:
$F^{B_s D^*_s}\approx F^{B D^*}$, $F^{B_s \phi}\approx F^{B \rho}\approx F^{B \omega}$,
 $F^{D_s \phi}\approx F^{D K^*}$ and $F^{D_s K^*}\approx F^{D \rho}\approx F^{D \omega}$,
 where $F$ represents any of the form factors, $V^{PV}, A^{PV}_0, A^{PV}_1$ or $A^{PV}_2$.
However, the slope parameters show considerable differences for these cases.

\item It is observed that heavy-to-light
form factors for the bottom mesons are smaller (between 0.2 to 0.4) than all
the charmed meson form factors and the heavy-to-heavy bottom meson form
factors, $F^{BD^*(B_sD^*_s)}$, which lie between 0.6 to 1. We also
notice the pattern $A^{PV}_0 > A^{PV}_1 > A^{PV}_2$ for all transitions,
whereas $V^{PV} > A^{PV}_0$ for charmed meson  and $B/B_s \to D/D_s$ transitions,
but $V^{PV}$ is slightly smaller than $A^{PV}_0$ and remains
 greater than $A^{PV}_{1,2}$ for the heavy-to-light bottom transitions.

\item Here also due to the reliability in fixing the $\beta$ parameter for lighter vector mesons
and the parent pseudoscalar mesons, the errors in the calculated form factors
$F(0)$ are very small. In contrast, the slope parameters do show
sensitivity to the variation in the $\beta$ parameters specially in the bottom sector.

\item Form factors,  $V^{PV}(0), A^{PV}_0(0)$, and $A^{PV}_1(0)$, for the
charm sector usually tend to decrease (increase) with increasing (decreasing) $\beta$
for initial meson, whereas they tend to increase (decrease) with increasing
 (decreasing) $\beta$ for final meson. However, the form factor $A^{PV}_2(0)$
 shows the opposite trend.

\item For the bottom sector, form factors generally tend to increase (decrease)
with increasing (decreasing) $\beta$ for each of the initial and the final mesons.

\item Slope parameters for all the cases are found to carry positive values. Both parameters
 ($a$ and $b$) generally
tend to increase (decrease) with decrease (increase) in $\beta$ for
each of the initial and final mesons.

\item The parameters $a$ is usually less sensitive to the $\beta$ variation,
whereas $b$ is more sensitive to $\beta$ values and may show large
variation $(10 \%)$ or even more sometimes for the bottom sector.

\item	Almost all the form factors for $D$ as well as $B$ are higher by $(5-10)\%$
than that obtained in the earlier work \cite{Cheng04}, whereas
both slope parameter are reduced in magnitude. This could happen because now we
perform 5-point fit for $q^2$ values.

\item On comparison with $ P \to P, V$ form factors obtained in the BSW model \cite{BSW}, the
Melikhov-Stech (MS) model \cite{Melikhov}, QCD sum rule (QSR) \cite{Ball91},
light-cone sum rules (LCSR) \cite{LCSR} and lattice calculations \cite{LatticeFF},
it is pointed out that our predictions agree well with the available lattice
results, and are most close to that of the MS model except for $B_s$ transitions,
which larger than our results. The LCSR and BSW model results are usually larger
for $P \to V$ form factors for $D$ and $B$ transitions, however
LCSR form factors for $B_s \to K^*$ transition match well with present work. The QSR calculations
are generally lower than our results, except for $B \to K^*$ form factors which are
higher than our predictions. Recently, $P \to V $
form factors for bottom mesons have also been calculated
in the perturbative QCD approach \cite{RHLi09},
which are found to be lower than the values obtained in the present work.

\item Experimentally, the form factors ratios $r_V \equiv V^{PV}(0)/A^{PV}_1(0)$ and
$r_2 \equiv A^{PV}_2(0)/A^{PV}_1(0)$ are available for two semileptonic decays
$D \to K^* \ell \nu$ and $D \to \phi \ell \nu$ \cite{PDG10}:
\be \label{eq:DKstar}
 r_V(D \to K^*) = 1.62 \pm 0.08, ~~  r_2(D \to K^*) = 0.83 \pm 0.05,
 \non\\
 r_V(D_s \to \phi) = 1.82 \pm 0.08, ~~  r_2(D_s \to \phi) = 0.84 \pm 0.11.
 \en
Our predictions $r_2(D \to K^*) = 0.83$ and $r_2(D_s \to \phi)= 0.86$,
agree well for both the decays, whereas $r_V(D \to K^*) = 1.36$
and $r_V(D_s \to \phi)= 1.42$ are lower than the corresponding experimental values.

\end{itemize}

\begin{table}[b]
\caption{Form factors of
$P(0^-)\rightarrow V(1^-)$ transitions
obtained in the covariant light-front model are fitted to the
3-parameter form Eq. (\ref{eq:FFpara}). All the form factors are dimensionless. }
 \label{tab:P2V}
\begin{ruledtabular}
\begin{tabular}{| c c c c | c c c c |}
$F$
    & $F(0)$
    & $a$
    & $b$
    & $F$
    & $F(0)$
    & $a$
    & $b$
    \\
\hline
$V^{D  \rho}$
    & $0.88^{-0.02+0.01}_{+0.01-0.01}$
    & $1.23^{-0.01-0.00}_{+0.01+0.00}$
    & $0.40^{-0.03-0.01}_{+0.04+0.01}$
    & $A^{D  \rho}_0$
    & $0.69^{-0.01+0.01}_{+0.01-0.01}$
    & $1.08^{-0.02-0.00}_{+0.01+0.00}$
    & $0.45^{-0.03-0.01}_{+0.03+0.01}$
    \\
$A^{D  \rho}_1$
    & $0.60^{-0.00+0.00}_{+0.00-0.01}$
    & $0.46^{-0.02-0.01}_{+0.02+0.01}$
    & $0.01^{-0.00-0.00}_{+0.00+0.00}$
    & $A^{D  \rho}_2$
    & $0.47^{+0.00-0.00}_{-0.00+0.00}$
    & $0.89^{+0.00-0.02}_{-0.00+0.02}$
    & $0.23^{-0.02-0.01}_{+0.02+0.01}$
    \\
$V^{D  \omega}$
    & $0.85^{-0.02+0.01}_{+0.01-0.01}$
    & $1.24^{-0.01-0.00}_{+0.01+0.00}$
    & $0.45^{-0.04-0.01}_{+0.04+0.01}$
    & $A^{D  \omega}_0$
    & $0.64^{-0.01+0.01}_{+0.01-0.01}$
    & $1.08^{-0.02+0.00}_{+0.01-0.00}$
    & $0.50^{-0.04-0.01}_{+0.04+0.01}$
    \\
$A^{D  \omega}_1$
    & $0.58^{-0.01+0.00}_{+0.00-0.00}$
    & $0.49^{-0.02-0.01}_{+0.02+0.01}$
    & $0.02^{-0.00-0.00}_{+0.01+0.00}$
    & $A^{D  \omega}_2$
    & $0.49^{+0.00-0.00}_{-0.00+0.00}$
    & $0.95^{-0.00-0.01}_{-0.00+0.01}$
    & $0.28^{-0.02-0.01}_{+0.02+0.01}$
    \\
$V^{D K^*}$
    & $0.98^{-0.02+0.01}_{+0.02-0.01}$
    & $1.10^{-0.02-0.00}_{+0.02+0.00}$
    & $0.32^{-0.03-0.01}_{+0.03+0.01}$
    & $A^{D K^*}_0$
    & $0.78^{-0.01+0.01}_{+0.01-0.01}$
    & $1.01^{-0.02-0.00}_{+0.02+0.00}$
    & $0.34^{-0.03-0.01}_{+0.03+0.01}$
    \\
$A^{D K^*}_1$
    & $0.72^{-0.01+0.01}_{+0.01-0.01}$
    & $0.45^{-0.02-0.01}_{+0.02+0.01}$
    & $0.01^{-0.00-0.00}_{+0.00+0.00}$
    & $A^{D K^*}_2$
    & $0.60^{+0.00-0.00}_{-0.00+0.00}$
    & $0.89^{-0.01-0.01}_{+0.00+0.01}$
    & $0.21^{-0.02-0.01}_{+0.02+0.01}$
    \\
\hline
$V^{D_s  K^*}$
    & $0.87^{-0.01+0.01}_{+0.01-0.01}$
    & $1.13^{+0.00+0.00}_{-0.01-0.00}$
    & $0.69^{-0.04-0.02}_{+0.05+0.03}$
    & $A^{D_s  K^*}_0$
    & $0.61^{-0.00+0.01}_{+0.00-0.01}$
    & $0.90^{+0.01+0.01}_{-0.01+0.02}$
    & $0.87^{-0.04-0.03}_{+0.05+0.01}$
    \\
$A^{D_s  K^*}_1$
    & $0.56^{-0.00+0.01}_{+0.00-0.01}$
    & $0.59^{-0.01-0.01}_{+0.01+0.01}$
    & $0.08^{-0.01-0.00}_{+0.01+0.01}$
    & $A^{D_s  K^*}_2$
    & $0.46^{+0.00-0.00}_{-0.00+0.00}$
    & $0.90^{+0.01-0.01}_{-0.01+0.01}$
    & $0.43^{-0.02-0.02}_{+0.02+0.02}$
    \\
$V^{D_s  \phi}$
    & $0.98^{-0.01+0.00}_{+0.01-0.00}$
    & $1.04^{-0.00+0.00}_{+0.00-0.00}$
    & $0.54^{-0.03-0.00}_{+0.04+0.01}$
    & $A^{D_s  \phi}_0$
    & $0.72^{-0.01+0.00}_{+0.01-0.00}$
    & $0.92^{-0.00+0.00}_{-0.00-0.00}$
    & $0.62^{-0.03-0.00}_{+0.04+0.01}$
    \\
$A^{D_s \phi}_1$
    & $0.69^{-0.00+0.00}_{+0.00-0.00}$
    & $0.56^{-0.02-0.00}_{+0.02+0.00}$
    & $0.07^{-0.01-0.00}_{+0.01+0.00}$
    & $A^{D_s  \phi}_2$
    & $0.59^{+0.00-0.00}_{-0.00+0.00}$
    & $0.90^{-0.00-0.00}_{-0.00+0.00}$
    & $0.38^{-0.02-0.00}_{+0.02+0.00}$
    \\
\hline
$V^{B  \rho}$
    & $0.29^{-0.00+0.01}_{-0.00-0.01}$
    & $1.77^{-0.03-0.01}_{+0.03+0.01}$
    & $1.06^{-0.06-0.03}_{+0.07+0.03}$
    & $A^{B  \rho}_0$
    & $0.32^{+0.00+0.01}_{-0.00-0.01}$
    & $1.67^{-0.03-0.01}_{+0.03+0.03}$
    & $1.01^{-0.04-0.02}_{+0.05-0.02}$
    \\
$A^{B  \rho}_1$
    & $0.24^{+0.00+0.00}_{-0.00-0.00}$
    & $0.86^{-0.03-0.01}_{+0.03+0.01}$
    & $0.15^{-0.02-0.01}_{+0.02+0.01}$
    & $A^{B  \rho}_2$
    & $0.22^{+0.00+0.00}_{-0.00-0.00}$
    & $1.56^{-0.02-0.02}_{+0.02+0.02}$
    & $0.85^{-0.05-0.03}_{+0.05+0.03}$
    \\
$V^{B  \omega}$
    & $0.27^{+0.00+0.00}_{-0.00-0.00}$
    & $1.81^{-0.03-0.01}_{+0.03+0.01}$
    & $1.18^{-0.07-0.02}_{+0.08+0.02}$
    & $A^{B  \omega}_0$
    & $0.28^{+0.00+0.01}_{-0.00-0.01}$
    & $1.62^{+0.05+0.07}_{+0.11+0.09}$
    & $1.22^{-0.15-0.11}_{-0.03-0.08}$
    \\
$A^{B  \omega}_1$
    & $0.23^{+0.00+0.00}_{-0.00-0.00}$
    & $0.91^{-0.03-0.01}_{+0.03+0.01}$
    & $0.18^{-0.02-0.01}_{+0.02+0.01}$
    & $A^{B  \omega}_2$
    & $0.21^{+0.00+0.00}_{-0.00-0.00}$
    & $1.62^{-0.03-0.01}_{+0.03+0.01}$
    & $0.97^{-0.06-0.03}_{+0.06+0.02}$
    \\
$V^{B  K^*}$
    & $0.36^{-0.00+0.01}_{+0.00-0.01}$
    & $1.69^{-0.03-0.01}_{+0.03+0.01}$
    & $0.95^{-0.06-0.02}_{+0.06+0.02}$
    & $A^{B K^*}_0$
    & $0.38^{+0.00+0.01}_{-0.00-0.01}$
    & $1.61^{-0.03-0.01}_{+0.03+0.01}$
    & $0.89^{-0.04-0.02}_{+0.05+0.02}$
    \\
$A^{B K^*}_1$
    & $0.31^{+0.00+0.00}_{-0.00-0.00}$
    & $0.84^{-0.03-0.01}_{+0.03+0.01}$
    & $0.12^{-0.02-0.01}_{+0.02+0.01}$
    & $A^{B K^*}_2$
    & $0.28^{+0.00+0.00}_{-0.00-0.00}$
    & $1.53^{-0.02-0.01}_{+0.02+0.01}$
    & $0.79^{-0.04-0.02}_{+0.05+0.02}$
    \\
$V^{B D^*}$
    & $0.77^{-0.01+0.02}_{+0.00-0.03}$
    & $1.25^{-0.02-0.03}_{+0.02+0.02}$
    & $0.38^{-0.02-0.03}_{+0.02+0.03}$
    & $A^{B D^*}_0$
    & $0.68^{-0.00+0.04}_{+0.00-0.04}$
    & $1.21^{-0.02-0.03}_{+0.02+0.02}$
    & $0.36^{-0.02-0.03}_{+0.02+0.03}$
    \\
$A^{B D^*}_1$
    & $0.65^{-0.00+0.02}_{+0.00-0.02}$
    & $0.60^{-0.01-0.03}_{+0.01+0.02}$
    & $0.00^{-0.00-0.01}_{+0.01+0.01}$
    & $A^{B D^*}_2$
    & $0.61^{-0.00-0.01}_{-0.00-0.00}$
    & $1.12^{-0.01-0.05}_{+0.01+0.04}$
    & $0.31^{-0.01-0.04}_{+0.01+0.04}$
    \\
\hline
$V^{B_s  K^*}$
    & $0.23^{+0.00+0.01}_{-0.00-0.01}$
    & $2.03^{-0.04-0.01}_{+0.04+0.01}$
    & $2.27^{-0.20-0.07}_{+0.22+0.08}$
    & $A^{B_s  K^*}_0$
    & $0.25^{+0.00+0.01}_{-0.00-0.01}$
    & $1.95^{-0.04-0.01}_{+0.04+0.01}$
    & $2.20^{-0.16-0.07}_{+0.18+0.08}$
    \\
$A^{B_s  K^*}_1$
    & $0.19^{+0.00+0.00}_{-0.00-0.01}$
    & $1.24^{-0.05-0.02}_{+0.05+0.02}$
    & $0.62^{-0.07-0.03}_{+0.09+0.03}$
    & $A^{B_s  K^*}_2$
    & $0.16^{+0.00+0.00}_{-0.00-0.00}$
    & $1.83^{-0.04-0.02}_{+0.04+0.02}$
    & $1.85^{-0.15-0.07}_{+0.17+0.08}$
    \\
$V^{B_s  \phi}$
    & $0.29^{+0.00+0.00}_{-0.00-0.00}$
    & $1.95^{-0.04-0.00}_{+0.04+0.00}$
    & $1.98^{-0.17-0.02}_{+0.19+0.02}$
    & $A^{B_s \phi}_0$
    & $0.31^{+0.00+0.00}_{-0.00-0.00}$
    & $1.87^{+0.02-0.00}_{+0.04+0.00}$
    & $1.87^{-0.31-0.02}_{+0.16+0.02}$
    \\
$A^{B_s  \phi}_1$
    & $0.25^{+0.00+0.00}_{-0.01-0.00}$
    & $1.20^{-0.05-0.01}_{+0.05+0.01}$
    & $0.54^{-0.06-0.01}_{+0.07+0.01}$
    & $A^{B_s  \phi}_2$
    & $0.22^{+0.00+0.00}_{-0.01-0.00}$
    & $1.79^{-0.04-0.00}_{+0.04+0.00}$
    & $1.67^{-0.13-0.02}_{+0.15+0.02}$
    \\
$V^{B_s D^*_s}$
    & $0.75^{-0.00+0.03}_{+0.00-0.04}$
    & $1.37^{-0.03-0.05}_{+0.03+0.04}$
    & $0.67^{-0.05-0.08}_{+0.05+0.10}$
    & $A^{B_s D^*_s}_0$
    & $0.66^{-0.00+0.04}_{-0.00-0.05}$
    & $1.33^{-0.03-0.04}_{+0.03+0.04}$
    & $0.63^{-0.04-0.08}_{+0.05+0.10}$
    \\
$A^{B_s D^*_s}_1$
    & $0.62^{+0.00+0.02}_{-0.00-0.03}$
    & $0.76^{-0.03-0.05}_{+0.03+0.05}$
    & $0.13^{-0.02-0.03}_{+0.02+0.04}$
    & $A^{B_s D^*_s}_2$
    & $0.57^{+0.00+0.00}_{-0.00-0.01}$
    & $1.25^{-0.02-0.07}_{+0.02+0.07}$
    & $0.56^{-0.04-0.09}_{+0.04+0.11}$
    \\
\end{tabular}
\end{ruledtabular}
\end{table}

\subsection{ $P(0^-) \to S (0^+)$  Form Factors}
\begin{itemize}

\item It has been discussed in Sec. III that there are two sets of scalar mesons: the
light scalar nonet formed by $\sigma(600)$,
$\kappa(800)$, $f_0(980)$ and $a_0(980)$; and the heavy scalar nonet
contains $a_0(1450)$, $K_0^*(1430)$, $f_0(1370)$ and
$f_0(1500)/f_0(1710)$. Though their underlying quark structure
is still controversial, the present experimental data seem
to provide a consistent picture that light scalar mesons
below or near 1 GeV can be described by the $qq \overline{q} \overline{q}$ states,
while scalars above 1 GeV form a conventional $q \overline{q}$ with possible
mixing with glueball states. In this work, we have
calculated the form factors involving heavy scalar mesons,
taking $f_0(1710)$ to be primarily a glueball, and $f_0(1500)$ and $f_0(1370)$
to be the SU(3) states as described in Sec. III.

\item From Table~\ref{tab:P2S}, we notice that all the form factors
for charmed mesons are around 0.5-0.6 where as all the
bottom meson form factors lie 0.25 to  0.30, and thus are roughly half of the
charmed meson form factors. Particularly, we note the following patterns:
$F^{B_s K^*_0} = F^{B a_0} = F^{B f_{0q}}$,
$F^{B_s f_{0s}} \approx F^{B D^*_0} = F^{B K^*_0}$, and
$F^{D_s f_{0s}}\approx F^{D a_0} = F^{D f_{0q}}$, where
$f_{0q}= (u\overline{u} + d\overline{d})/\sqrt{2}$, and
$f_{0s}$ is a pure $(s\overline{s})$ state. Here too, the slope parameters
show considerable differences among for the related form factors.

\item Both of the form factors $F^{PS}_{1,0}(0)$ decrease (increase)
 with increasing (decreasing) $\beta$ for initial meson,
 whereas they increase (decrease) with increasing (decreasing) $\beta$ for final meson.

\item The slope parameters $b$ for both the form factors $F^{PS}_{1,0}$, and
$a$ for $F^{PS}_1$ are found to be positive. For $F^{PS}_0$ form factor, $a$ turns out to be negative
when charmed mesons appear in either initial or final state.

\item For the bottom sector, the slope parameters are larger than that for the charm sector.
The parameter $b$ is generally much small ($< 0.1$) for $F^{PS}_0(0)$, except for
$F^{B_s \to S(0^+)}_0$ and $F^{BD^*}_0$, for which $b$ could be as big as 0.45.

\item	For transitions of the charmed mesons, the slope parameters
for the form factor $F^{PS}_0$ are more sensitive to change
in $\beta$ for each of the initial and final mesons. However, these are
less sensitive for $F^{PS}_1$.

\item	Slope parameters (except for the case of negative $a$)
show an increase (decrease) with decrease (increase) in $\beta$ for each
of the initial and final mesons.

\item	No significant change is found in the form factors, though $a$ and $b$
are slightly lowered that their values obtained in the earlier work \cite{Cheng04}.
Based on the light-cone sum rules, Chernyak \cite{Chernyak} has
estimated the $F^{B a_0(1450)}_{1,0}(0)$ =0.46, while our result is 0.25
and is similar to the $B\to \pi$ form factor at $q^2=0$.

\item On comparison of the $P \to S $ and $P \to P $ form factors,
we notice $F^{D\to S} < F^{D \to P}$ for the same flavor
content of the final state mesons. For the bottom sector,
$F^{B D^*_0} < F^{B D}$ and
$F^{B_s D^*_{s0}} < F^{B_s D_s}$, for heavy-to-heavy transitions, while
$F^{B \to S} \approx F^{B \to P}$ for heavy-to-light transitions.
It has been pointed out before \cite{Cheng04} that the suppression of
the $ B \to D^*_0$ form factor relative to that of $B \to D$ is supported by experiment.

\item An inspection of Table~\ref{tab:P2S}
indicates that similar to the $P \to P $ transitions, the
form factors $F^{PS}_0$ generally show a monopole behavior, and $F^{PS}_1$
have a dipole behavior particularly for charmed meson
transitions. In general, form factors for $P \to S $
transitions increase slowly with $q^2$ compared to that for $P \to P $ ones.

\end {itemize}

{\squeezetable
\begin{table}[b]
\caption{Form factors of
$P(0^-)\rightarrow S(0^+)$ transitions
obtained in the covariant light-front model are fitted to the
3-parameter form Eq. (\ref{eq:FFpara}). All the form factors are dimensionless. }
 \label{tab:P2S}
\begin{ruledtabular}
\begin{tabular}{|cccc|cccc|}
$F$
    & $F(0)$
    & $a$
    & $b$
    & $F$
    & $F(0)$
    & $a$
    & $b$
    \\
\hline
$F^{D  a_0}_1$
    & $0.51^{-0.01+0.01}_{+0.01-0.02}$
    & $1.06^{-0.02-0.02}_{+0.01+0.01}$
    & $0.24^{-0.02-0.02}_{+0.02+0.02}$
    & $F^{D  a_0}_0$
    & $0.51^{-0.01+0.02}_{+0.01-0.02}$
    & $-0.04^{+0.07-0.06}_{-0.06+0.07}$
    & $0.02^{-0.02+0.01}_{+0.02-0.02}$
    \\
$F^{D f_{0q}}_1$
    & $0.51^{-0.01+0.03}_{+0.01-0.04}$
    & $1.06^{-0.02-0.04}_{+0.01+0.03}$
    & $0.24^{-0.02-0.04}_{+0.02+0.04}$
    & $F^{D  f_{0q}}_0$
    & $0.51^{-0.01+0.04}_{+0.01-0.05}$
    & $-0.04^{+0.07-0.13}_{-0.06+0.19}$
    & $0.02^{-0.02+0.03}_{+0.02-0.04}$
    \\
$F^{DK^*_0}_1$
    & $0.47^{-0.01+0.02}_{+0.01-0.02}$
    & $0.94^{-0.02-0.01}_{+0.02+0.01}$
    & $0.19^{-0.02-0.01}_{+0.02+0.02}$
    & $F^{DK^*_0}_0$
    & $0.47^{-0.01+0.02}_{+0.01-0.03}$
    & $-0.31^{+0.04-0.04}_{-0.04+0.05}$
    & $0.08^{-0.01+0.01}_{+0.01-0.01}$
    \\
\hline
$F^{D_s K^*_0}_1$
    & $0.55^{-0.01+0.02}_{+0.01-0.03}$
    & $1.02^{-0.01-0.01}_{+0.00+0.01}$
    & $0.38^{-0.02-0.04}_{+0.02+0.06}$
    & $F^{D_s K^*_0}_0$
    & $0.55^{-0.01+0.02}_{+0.01-0.03}$
    & $-0.04^{+0.02-0.05}_{-0.02+0.07}$
    & $0.05^{-0.01+0.01}_{+0.01-0.01}$
    \\
$F^{D_s f_{0s}}_1$
    & $0.52^{-0.01+0.01}_{+0.01-0.01}$
    & $0.91^{-0.01-0.00}_{+0.01+0.00}$
    & $0.29^{-0.02-0.01}_{+0.02+0.01}$
    & $F^{D_s f_{0s}}_0$
    & $0.52^{-0.01+0.01}_{+0.01-0.01}$
    & $-0.34^{+0.01-0.01}_{-0.01+0.01}$
    & $0.10^{-0.01+0.00}_{+0.01-0.00}$
    \\
\hline
$F^{B a_0}_1$
    & $0.25^{-0.00+0.01}_{+0.00-0.01}$
    & $1.53^{-0.03-0.01}_{+0.03+0.01}$
    & $0.64^{-0.04-0.04}_{+0.05+0.04}$
    & $F^{B a_0}_0$
    & $0.25^{-0.00+0.01}_{+0.00-0.01}$
    & $0.54^{-0.01-0.03}_{+0.01+0.03}$
    & $0.01^{-0.01-0.00}_{+0.02+0.00}$
    \\
$F^{B f_{0q}}_1$
    & $0.25^{-0.00+0.03}_{+0.00-0.03}$
    & $1.53^{-0.03-0.03}_{+0.03+0.03}$
    & $0.64^{-0.04-0.08}_{+0.05+0.11}$
    & $F^{B f_{0q}}_0$
    & $0.25^{-0.00+0.03}_{+0.00-0.03}$
    & $0.54^{-0.01-0.06}_{+0.01+0.07}$
    & $0.01^{-0.01-0.00}_{+0.02+0.01}$
    \\
$F^{BK^*_0}_1$
    & $0.27^{-0.01+0.01}_{+0.01-0.02}$
    & $1.43^{-0.03-0.01}_{+0.03+0.01}$
    & $0.52^{-0.04-0.03}_{+0.04+0.04}$
    & $F^{BK^*_0}_0$
    & $0.27^{-0.01+0.01}_{+0.01-0.02}$
    & $0.32^{-0.01-0.02}_{+0.01+0.03}$
    & $0.05^{-0.01+0.00}_{+0.01+0.00}$
    \\
$F^{BD^*_0}_1$
    & $0.27^{-0.01+0.03}_{+0.01-0.03}$
    & $1.08^{-0.04+0.03}_{+0.04-0.07}$
    & $0.23^{-0.02-0.00}_{+0.02+0.00}$
    & $F^{B D^*_0}_0$
    & $0.27^{-0.01+0.03}_{+0.01-0.03}$
    & $-0.48^{+0.02-0.03}_{-0.02+0.01}$
    & $0.36^{-0.03+0.03}_{+0.03-0.03}$
    \\
\hline
$F^{B_s K^*_0}_1$
    & $0.25^{-0.00+0.02}_{+0.00-0.02}$
    & $1.75^{-0.04-0.04}_{+0.04+0.05}$
    & $1.33^{-0.12-0.12}_{+0.14+0.18}$
    & $F^{B_s K^*_0}_0$
    & $0.25^{-0.00+0.02}_{+0.00-0.02}$
    & $0.74^{-0.03-0.06}_{+0.03+0.07}$
    & $0.24^{-0.04-0.03}_{+0.04+0.05}$
    \\
$F^{B_s f_{0s}}_1$
    & $0.28^{-0.00+0.01}_{+0.00-0.01}$
    & $1.64^{-0.04-0.01}_{+0.04+0.01}$
    & $1.07^{-0.10-0.04}_{+0.11+0.04}$
    & $F^{B_s f_{0s}}_0$
    & $0.28^{-0.00+0.01}_{+0.00-0.01}$
    & $0.52^{-0.03-0.02}_{+0.03+0.02}$
    & $0.20^{-0.03-0.01}_{+0.03+0.01}$
    \\
$F^{B_s D^*_{0s}}_1$
    & $0.30^{-0.02+0.03}_{+0.02-0.03}$
    & $1.18^{-0.06+0.01}_{+0.06-0.04}$
    & $0.51^{-0.05-0.05}_{+0.06+0.05}$
    & $F^{B_s D^*_{0s}}_0$
    & $0.30^{-0.02+0.03}_{+0.02-0.04}$
    & $-0.47^{-0.01-0.02}_{+0.01+0.01}$
    & $0.45^{-0.02+0.01}_{+0.02+0.00}$
     \\
\end{tabular}
\end{ruledtabular}
\end{table}
}

\subsection{$P(0^-) \to A (1^+\,:~ ^3P_1)$  Form Factors}

\begin{itemize}

\item In Tables~\ref{tab:P2Ap} and \ref{tab:P2An}, we have given heavy-to-light
form factors involving axial vector nonet mesons, whereas
heavy to heavy form factors are separately presented in
Table ~\ref{tab:B2D} with the final state charmed mesons
being taken as the heavy quark spin basis.

\item From Table~\ref{tab:P2Ap}, all the form factors are found to be positive
for the bottom as well as charm sectors. We also notice the following pattern:
 $V^{PA}_1 > V^{PA}_0 > A^{PA} > V^{PA}_2$ for the charmed mesons
 and $V^{PA}_1 > A^{PA} > V^{PA}_2 > V^{PA}_0$ for the bottom mesons.

\item  Numerically speaking, the form factor $A^{PA}(0)$ for the bottom transitions
is generally around 0.25, and it is larger
than that for charmed meson transition for which it lies close 0.16. Similar
behavior is observed for $V^{PA}_2(0)$, which is $<0.1$ for the charm sector, where as
it lies around 0.2 for the bottom transitions. In contrast, the form factor
$V^{PA}_1(0)$, lying around 0.4 for the bottom sector, is significantly smaller than that
 for the charm sector, where its value lies between 1.4 to 1.8. Also $V^{PA}_0(0)$ for the
 bottom transitions is roughly half of its value for charmed meson
 transitions.

\item	The form factors are not very sensitive to the variation chosen for
the $\beta$ parameters. However, they generally
tend to decrease (increase) with increasing (decreasing) $\beta$ for initial meson, whereas
they increase (decrease) with increasing (decreasing) $\beta$ for final meson.

\item All the slope parameters, except for $V^{PA}_1$ and $V^{PA}_2$ for the charmed meson
 transitions, are found to be positive as per the definition given in
Eq.~(\ref{eq:FFpara}). For the bottom sector, the slope parameters
are significantly larger than that for the charm sector.

\item Slope parameters $a$  and $b$ for $A^{PA}$, and $V^{PA}_1$ are
generally less sensitive to $\beta$ variation, but for $V^{PA}_0$ and $V^{PA}_2$  they
could show more sensitivity (even up to $20\%$) to change in $\beta$ values.

\item	 Generally no large change occurs in the form factors  obtained in
the earlier work \cite{Cheng04}, though slope
parameters often show some changes.\footnote{Form factor  $A^{D K_1(^3P_1)}(0) = 0.98$ given
in the earlier \cite{Cheng04} is erroneous, and should be replaced
with 0.15.} This could happen since we now perform 5-point fit for $q^2$ values.
We find that  the form factors $V^{PA}_0$
 for $B \to a_1$  transition has marginally increased to 0.14.

\item There are several existing model calculations for
  $B\rightarrow A$ form factors: the ISGW2 model \cite{ISGW2},
  the constituent quark-meson model (CQM) \cite{Deandrea},
 the QCD sum rules (QSR) \cite{Aliev}, light cone sum rules
 (LCSR) \cite{KCYang08}, and more recently the
 perturbative QCD (pQCD) approach \cite{RHLi09}. For the sake of comparison, results
 for $B \to a_1$  transition form factors are given in Table \ref{tab:B2a1} for these
 approaches, which show quite significant differences since these approaches differ in
  their treatment of dynamics of form factors. For
  example, $V^{Ba_1}_0 = 1.20$, obtained in the quark-meson model and 1.01 in the ISGW2 model,
   are larger than the values obtained in other approaches. If $a_1 (1260)$
   behaves as the scalar partner of the $\rho$ meson,  $V^{Ba_1}_0$ is expected to
   be similar to  $A^{B\rho}_0$, which is of order 0.3 at $q^2 = 0$. Therefore, it
    appears to us that a magnitude of order unity for  $V^{Ba_1}_0 $ as predicted
     by the ISGW2 model and CQM is very unlikely. In principle, the experimental measurements
 of $\overline{B}^0\rightarrow a^{\pm}_1 \pi^{\mp}$ will enable us to test the
 form factors $V^{Ba_1}_0$. The BaBar and Belle measurements \cite{BABAR, Abe07} of
 $\overline{B}^0\rightarrow a^{\pm}_1 \pi^{\mp}$   favors a value of
$V_0^{Ba_1}(0)\approx 0.30$ \cite{ChengAP}, which is very close the LCSR result
shown in Table \ref{tab:B2a1}.

\end {itemize}

{\squeezetable
\begin{table}[b]
\caption{Form factors of
$P(0^-)\rightarrow A(1^{++})$ transitions
obtained in the covariant light-front model are fitted to the
3-parameter form Eq. (\ref{eq:FFpara}). All the form factors are dimensionless. }
 \label{tab:P2Ap}
\begin{ruledtabular}
\begin{tabular}{| c c c c | c c c c |}
$F$
    & $F(0)$
    & $a$
    & $b$
    & $F$
    & $F(0)$
    & $a$
    & $b$
    \\
\hline
$A^{D  a_1}$
    & $0.19^{-0.01+0.00}_{+0.01-0.01}$
    & $1.03^{-0.03-0.01}_{+0.02+0.00}$
    & $0.16^{-0.02-0.01}_{+0.02+0.01}$
    & $V^{D a_1}_0$
    & $0.32^{-0.00-0.00}_{+0.00-0.00}$
    & $0.96^{-0.01-0.01}_{+0.00+0.00}$
    & $0.43^{-0.06+0.01}_{+0.07-0.01}$
    \\
$V^{D  a_1}_1$
    & $1.51^{-0.04+0.00}_{+0.04-0.01}$
    & $-0.06^{-0.01-0.02}_{+0.01+0.02}$
    & $0.04^{-0.00+0.00}_{+0.00-0.00}$
    & $V^{D a_1}_2$
    & $0.05^{-0.01+0.00}_{+0.01-0.00}$
    & $-0.02^{-0.08-0.00}_{+0.07+0.00}$
    & $0.12^{+0.00-0.01}_{-0.00+0.01}$
    \\
$A^{D  f_{1q}}$
    & $0.18^{-0.01+0.01}_{+0.01-0.01}$
    & $1.03^{-0.03-0.02}_{+0.02+0.00}$
    & $0.16^{-0.02-0.02}_{+0.02+0.03}$
    & $V^{D f_{1q}}_0$
    & $0.34^{-0.00-0.00}_{+0.00-0.00}$
    & $0.97^{-0.02-0.01}_{+0.01-0.00}$
    & $0.39^{-0.05+0.03}_{+0.06+0.00}$
    \\
$V^{D  f_{1q}}_1$
    & $1.75^{-0.05-0.00}_{+0.04-0.02}$
    & $-0.02^{-0.01-0.05}_{+0.01+0.06}$
    & $0.04^{-0.00+0.01}_{+0.00-0.00}$
    & $V^{D f_{1q}}_2$
    & $0.05^{-0.01+0.00}_{+0.01-0.00}$
    & $-0.02^{-0.08-0.01}_{+0.07+0.01}$
    & $0.12^{+0.00-0.01}_{-0.00+0.02}$
    \\
$A^{D K_{1A}}$
    & $0.15^{-0.01+0.01}_{+0.01-0.01}$
    & $0.89^{-0.03+0.00}_{+0.03-0.01}$
    & $0.12^{-0.02-0.01}_{+0.02+0.01}$
    & $V^{D K_{1A}}_0$
    & $0.28^{-0.00-0.00}_{+0.00+0.00}$
    & $0.84^{-0.02-0.01}_{+0.01-0.01}$
    & $0.39^{-0.05+0.04}_{+0.06-0.03}$
    \\
$V^{D K_{1A}}_1$
    & $1.60^{-0.05-0.02}_{+0.05+0.01}$
    & $-0.22^{-0.00-0.03}_{+0.00+0.03}$
    & $0.07^{-0.00+0.00}_{+0.00-0.00}$
    & $V^{D K_{1A}}_2$
    & $0.01^{-0.00+0.00}_{+0.00-0.00}$
    & $-0.83^{-0.17+0.02}_{+0.15-0.03}$
    & $0.24^{+0.04-0.01}_{-0.03+0.01}$
    \\
\hline
$A^{D_s K_{1A}}$
    & $0.19^{-0.01+0.01}_{+0.01-0.01}$
    & $0.99^{-0.01-0.01}_{+0.01+0.01}$
    & $0.28^{-0.02-0.03}_{+0.02+0.04}$
    & $V^{D_s K_{1A}}_0$
    & $0.29^{-0.00-0.00}_{+0.00+0.00}$
    & $0.72^{+0.05-0.09}_{-0.06+0.07}$
    & $0.87^{-0.09+0.08}_{+0.10-0.05}$
    \\
$V^{D_s K_{1A}}_1$
    & $1.68^{-0.03-0.01}_{+0.03+0.00}$
    & $-0.04^{-0.01-0.04}_{+0.01+0.05}$
    & $0.06^{-0.00+0.00}_{+0.00-0.00}$
    & $V^{D_s K_{1A}}_2$
    & $0.07^{-0.00-0.00}_{+0.00-0.00}$
    & $0.22^{-0.03-0.01}_{+0.03+0.02}$
    & $0.15^{-0.01-0.01}_{+0.01+0.02}$
    \\
$A^{D_s  f_{1s}}$
    & $0.17^{-0.01+0.00}_{+0.01-0.00}$
    & $0.86^{-0.02-0.00}_{+0.02-0.00}$
    & $0.20^{-0.02-0.01}_{+0.02+0.01}$
    & $V^{D_s  f_{1s}}_0$
    & $0.22^{+0.00-0.00}_{-0.00+0.00}$
    & $0.19^{+0.13-0.13}_{-0.17+0.10}$
    & $1.20^{-0.15+0.11}_{+0.18-0.09}$
    \\
$V^{D_s  f_{1s}}_1$
    & $1.47^{-0.03-0.01}_{+0.03+0.01}$
    & $-0.29^{-0.01-0.01}_{+0.01+0.01}$
    & $0.09^{-0.00+0.00}_{+0.00-0.00}$
    & $V^{D_s  f_{1s}}_2$
    & $0.03^{-0.00+0.00}_{+0.00-0.00}$
    & $-0.34^{-0.05+0.00}_{+0.05-0.00}$
    & $0.17^{+0.00-0.00}_{-0.00+0.00}$
    \\
\hline
$A^{B  a_1}$
    & $0.24^{-0.01+0.01}_{+0.01-0.01}$
    & $1.48^{-0.03-0.01}_{+0.03+0.01}$
    & $0.57^{-0.04-0.03}_{+0.05+0.04}$
    & $V^{B a_1}_0$
    & $0.14^{+0.01+0.01}_{-0.01-0.01}$
    & $1.66^{-0.04-0.01}_{+0.04+0.01}$
    & $1.11^{-0.08-0.03}_{+0.09+0.02}$
    \\
$V^{B a_1}_1$
    & $0.36^{-0.01+0.01}_{+0.01-0.01}$
    & $0.26^{-0.02-0.02}_{+0.02+0.03}$
    & $0.14^{-0.01-0.00}_{+0.01+0.01}$
    & $V^{B a_1}_2$
    & $0.17^{-0.01+0.01}_{+0.01-0.01}$
    & $1.08^{-0.05-0.02}_{+0.05+0.02}$
    & $0.44^{-0.03-0.03}_{+0.09+0.04}$
    \\
$A^{B  f_{1q}}$
    & $0.24^{-0.01+0.02}_{+0.01-0.03}$
    & $1.48^{-0.03-0.03}_{+0.03+0.03}$
    & $0.57^{-0.04-0.08}_{+0.05+0.10}$
    & $V^{B f_{1q}}_0$
    & $0.14^{+0.01+0.01}_{-0.01-0.02}$
    & $1.65^{-0.04-0.02}_{+0.04+0.02}$
    & $1.07^{-0.08-0.06}_{+0.09+0.09}$
    \\
$V^{B f_{1q}}_1$
    & $0.37^{-0.01+0.03}_{+0.01-0.03}$
    & $0.27^{-0.02-0.05}_{+0.02+0.06}$
    & $0.13^{-0.01-0.01}_{+0.01+0.02}$
    & $V^{B f_{1q}}_2$
    & $0.17^{-0.01+0.01}_{+0.01-0.02}$
    & $1.08^{-0.05-0.04}_{+0.05+0.05}$
    & $0.44^{-0.03-0.07}_{+0.03+0.09}$
    \\
$A^{B K_{1A}}$
    & $0.27^{-0.01+0.01}_{+0.01-0.02}$
    & $1.39^{-0.04-0.01}_{+0.04+0.00}$
    & $0.47^{-0.04-0.03}_{+0.04+0.04}$
    & $V^{B K_{1A}}_0$
    & $0.16^{+0.01+0.01}_{-0.01-0.01}$
    & $1.55^{-0.03+0.01}_{+0.04-0.01}$
    & $1.00^{-0.10-0.02}_{+0.09+0.03}$
    \\
$V^{B K_{1A}}_1$
    & $0.39^{-0.01+0.01}_{+0.01-0.02}$
    & $0.07^{-0.02-0.02}_{+0.02+0.02}$
    & $0.19^{-0.00-0.00}_{+0.00+0.01}$
    & $V^{B K_{1A}}_2$
    & $0.17^{-0.01+0.01}_{+0.01-0.01}$
    & $0.84^{-0.06-0.01}_{+0.06+0.01}$
    & $0.36^{-0.02-0.03}_{+0.02+0.04}$
    \\
\hline
$A^{B_s K_{1A}}$
    & $0.24^{-0.00+0.02}_{+0.00-0.02}$
    & $1.70^{-0.05-0.04}_{+0.05+0.05}$
    & $1.22^{-0.12-0.12}_{+0.13+0.17}$
    & $V^{B_s K_{1A}}_0$
    & $0.12^{+0.01+0.01}_{-0.01-0.01}$
    & $1.88^{-0.05-0.02}_{+0.06+0.03}$
    & $2.06^{-0.25-0.08}_{+0.30+0.13}$
    \\
$V^{B_s K_{1A}}_1$
    & $0.37^{-0.00+0.02}_{+0.00-0.03}$
    & $0.53^{-0.04-0.06}_{+0.04+0.07}$
    & $0.29^{-0.03-0.03}_{+0.03+0.04}$
    & $V^{B_s K_{1A}}_2$
    & $0.17^{-0.00+0.01}_{+0.00-0.01}$
    & $1.40^{-0.07-0.04}_{+0.07+0.05}$
    & $0.97^{-0.10-0.10}_{+0.12+0.15}$
    \\
$A^{B_s  f_{1s}}$
    & $0.28^{-0.01+0.01}_{+0.01-0.01}$
    & $1.59^{-0.05-0.01}_{+0.05+0.01}$
    & $0.99^{-0.10-0.03}_{+0.11+0.04}$
    & $V^{B_s  f_{1s}}_0$
    & $0.13^{+0.01+0.00}_{-0.01-0.00}$
    & $1.79^{-0.06-0.00}_{+0.05+0.00}$
    & $2.00^{-0.25-0.01}_{+0.33+0.02}$
    \\
$V^{B_s  f_{1s}}_1$
    & $0.41^{-0.01+0.01}_{+0.01-0.01}$
    & $0.29^{-0.04-0.02}_{+0.04+0.02}$
    & $0.29^{-0.02-0.01}_{+0.02+0.01}$
    & $V^{B_s  f_{1s}}_2$
    & $0.18^{-0.01+0.00}_{+0.01-0.00}$
    & $1.18^{-0.07-0.01}_{+0.07+0.01}$
    & $0.74^{-0.07-0.03}_{+0.08+0.03}$
    \\
\end{tabular}
\end{ruledtabular}
\end{table}
}

{\squeezetable
\begin{table}[b1]
\caption{Form factors of
$B \rightarrow a_1$ transitions at maximum recoil ($q^2=0$). The results of CQM and
QSR have been rescaled according to the form factor definition in Eq. (\ref{eq:ffpdimless})}
 \label{tab:B2a1}
\begin{ruledtabular}
\begin{tabular}{| c c c c c c c |}
$B \rightarrow a_1$
    & $\textrm{This work}$
    & $\textrm{ISGW2}$ \cite{ISGW2}
    & $\textrm{CQM}$ \cite{Deandrea}
    & $\textrm{QSR}$ \cite{Aliev}
    & $\textrm{LCSR}$ \cite{KCYang08}
    & $\textrm{pQCD}$ \cite{RHLi09}
    \\
 \hline
$A$
 & $0.24^{-0.01+0.01}_{+0.01-0.01}$
 & $0.21$
 & $0.09$
 & $0.41\pm 0.06$
  & $0.48\pm 0.09$
 & $0.26^{+0.06+0.00+0.03}_{-0.05-0.01-0.03}$
     \\
$V_0$
   & $0.14^{+0.01+0.01}_{-0.01-0.01}$
 & $1.01$
 & $1.20$
 & $0.23\pm 0.05$
  & $0.30\pm 0.05$
 & $0.34^{+0.07+0.01+0.08}_{-0.07-0.02-0.08}$
    \\
$V_1$
    & $0.36^{-0.01+0.01}_{+0.01-0.01}$
 & $0.54$
 & $1.32$
 & $0.68\pm 0.08$
  & $0.37\pm 0.07$
 & $0.43^{+0.10+0.01+0.05}_{-0.09-0.01-0.05}$
    \\
$V_2$
    & $0.17^{-0.01+0.01}_{+0.01-0.01}$
 & $-0.05$
 & $0.34$
 & $0.33\pm 0.03$
  & $0.42\pm 0.08$
 & $0.13^{+0.03+0.00+0.00}_{-0.03-0.01-0.00}$
    \\
\end{tabular}
\end{ruledtabular}
\end{table}
}

\subsection{ $ P(0^-)\to A (1^+\,:~ ^1P_1 )$  Form Factors}

\begin{itemize}

\item From Table~\ref{tab:P2An}, we find that the form factors,
 $A^{PA}(0), V^{PA}_0(0)$ and $V^{PA}_1(0)$, are positive,
where as $V^{PA}_2(0)$ is negative and small (around -0.1) for the
bottom as well as charm sector, and follow the pattern:
 $V^{PA}_1 > V^{PA}_0 > A^{PA} > |V^{PA}_2| $ for the charmed mesons
 and  $V^{PA}_0 > V^{PA}_1 > A^{PA} \geq |V^{PA}_2| $  for the bottom mesons.
Numerically, form factors $A^{PA}(0)$ is generally
around 0.1, where as $V^{PA}_0(0)$ lies close to 0.5 for all the cases.
The form factor $V^{PA}_1(0)$, lying between 0.15 to 0.20, for the
bottom sector is significantly smaller than that
 for the charm sector, where its value lies between 1.3 to 1.6.

\item	The form factors
 $A^{PA}_0(0)$ and $V^{PA}_1(0)$ usually increase (decrease) with
 increasing (decreasing) beta for initial meson as well as for final meson.

\item The form factors $V^{PA}_0(0)$ and $V^{PA}_3(0)$ decrease (increase) in magnitude
with increasing (decreasing) $\beta$ for initial meson, and show the opposite
trend for final mesons, i.e., these increase (decrease) with
increasing (decreasing) $\beta$ for the final state.

\item All the slope parameters are found to be positive. For the bottom sector,
the slope parameters are larger than that for the charm sector.

\item Slope parameters $a$ and $b$ for $A^{PA}, V^{PA}_0$, and $V^{PA}_1$
are less sensitive (a few $\%$) to the variation in the $\beta$ values. For
$V^{PA}_2$ form factor, the slope parameters show huge sensitivity to the change in beta values,
even with assuming  $q^2$ behavior given by Eq. (\ref{eq:FFpara}).
However, these are less sensitive for $B \to b_1/h_{1q}$ cases.

\item	No significant change is found in the form factors obtained in
the earlier work \cite{Cheng04}, however, the slope parameters show
difference.

\item While comparing the form factors of heavy-to-light spin $1$ meson transitions,
we notice the following relations for the same flavor content of the mesons:
$A^{P A(1^{++})}(0) > A^{P A(1^{+-})}(0)~$,
$V^{P A(1^{++})}_1(0) > V^{P  A(1^{+-})}_1(0)$. But for $V^{PA}_0(0)$ form factors, we find
$V^{P A(1^{++})}_0(0) < V^{P  A(1^{+-})}_0(0)$. For $V^{PA}_2(0)$ form factors, we observe
opposite behavior for the charmed and bottom mesons, i.e.,
$|V^{D, D_s\to A(1^{++})}_2(0)| < |V^{D, Ds \to A(1^{+-})}_2(0)|$,
whereas $|V^{B, Bs \to A(1^{++})}_2(0)| > |V^{B, Bs \to A(1^{+-})}_2(0)|$.

\item While comparing the form factors of heavy-to-light vector and axial-vector mesons,
we notice the following patterns:
$V^{P V}(0) > A^{P A(1^{++})}(0)$ for the same flavor content of the mesons;
$ V^{P  A(1^{+-})}_1(0)> A^{P V}_1(0)$ for charmed mesons
and $A^{P V}_1(0) > V^{P  A(1^{+-})}_1(0)$ for the bottom mesons.
But for $V^{PA}_0(0)$ form factors, we find
$V^{P A(1^{+-})}_0(0)< A^{P V}_0(0)$ for charmed mesons
and $V^{P A(1^{++})}_0(0)< A^{P V}_0(0) < V^{P A(1^{+-})}_0(0)$ for the bottom mesons.
We also observe that $A^{P V}_2(0)$ is higher than both $V^{D, Ds \to A(1^{++})}_2(0)$
as well as $|V^{D, Ds \to A(1^{+-})}_2(0)| $.

\end {itemize}

{\squeezetable
\begin{table}[b]
\caption{Form factors of
$P(0^-)\rightarrow A(1^{+-})$ transitions
obtained in the covariant light-front model are fitted to the
3-parameter form Eq. (\ref{eq:FFpara}). All the form factors are dimensionless.}
 \label{tab:P2An}
\begin{ruledtabular}
\begin{tabular}{| c c c c | c c c c |}
$F$
    & $F(0)$
    & $a$
    & $b$
    & $F$
    & $F(0)$
    & $a$
    & $b$
    \\
\hline
$A^{D  b_1}$
    & $0.12^{+0.00+0.00}_{-0.00-0.02}$
    & $1.09^{-0.01-0.01}_{+0.01+0.01}$
    & $0.50^{-0.04-0.04}_{+0.04+0.04}$
    & $V^{D b_1}_0$
    & $0.50^{-0.01+0.02}_{+0.01-0.02}$
    & $0.98^{-0.02-0.01}_{+0.02-0.00}$
    & $0.26^{-0.00-0.02}_{+0.00+0.03}$
    \\
$V^{D  b_1}_1$
    & $1.39^{+0.02+0.03}_{-0.02-0.04}$
    & $0.44^{-0.03-0.02}_{+0.03+0.02}$
    & $0.05^{-0.01-0.00}_{+0.01+0.01}$
    & $V^{D b_1}_2$
    & $-0.10^{+0.02-0.01}_{-0.02+0.01}$
    & $0.26^{-0.64+0.13}_{+0.28-0.23}$
    & $0.90^{+0.46-0.16}_{-0.23+0.23}$
    \\
$A^{D  h_{1q}}$
    & $0.11^{+0.00+0.00}_{-0.00-0.01}$
    & $1.09^{-0.01-0.02}_{+0.01+0.01}$
    & $0.50^{-0.04-0.08}_{+0.04+0.10}$
    & $V^{D h_{1q}}_0$
    & $0.49^{-0.01+0.04}_{+0.01-0.05}$
    & $0.98^{-0.02-0.03}_{+0.02-0.02}$
    & $0.26^{-0.00-0.06}_{+0.00+0.06}$
    \\
$V^{D  h_{1q}}_1$
    & $1.42^{+0.02+0.06}_{-0.02-0.09}$
    & $0.44^{-0.03-0.04}_{+0.03+0.04}$
    & $0.05^{-0.01-0.01}_{+0.01+0.02}$
    & $V^{D h_{1q}}_2$
    & $-0.10^{+0.02-0.02}_{-0.01+0.03}$
    & $0.26^{-0.64+0.24}_{+0.28-0.79}$
    & $0.90^{+0.46-0.33}_{-0.23+0.69}$
    \\
$A^{D K_{1B}}$
    & $0.10^{+0.00+0.00}_{-0.00-0.00}$
    & $0.98^{-0.01-0.01}_{+0.01+0.01}$
    & $0.37^{-0.03-0.03}_{+0.03+0.04}$
    & $V^{D K_{1B}}_0$
    & $0.48^{-0.01+0.02}_{+0.01-0.03}$
    & $0.94^{-0.02-0.02}_{+0.01+0.01}$
    & $0.22^{+0.00-0.03}_{-0.00+0.03}$
    \\
$V^{D K_{1B}}_1$
    & $1.58^{+0.02+0.03}_{-0.03-0.05}$
    & $0.31^{-0.02-0.01}_{+0.02+0.02}$
    & $0.04^{-0.00-0.00}_{+0.00+0.01}$
    & $V^{D K_{1B}}_2$
    & $-0.13^{+0.01-0.01}_{-0.01+0.01}$
    & $0.57^{-0.06-0.01}_{+0.04-0.01}$
    & $0.32^{+0.05-0.04}_{-0.03+0.06}$
    \\
\hline
$A^{D_s K_{1B}}$
    & $0.10^{+0.00+0.00}_{-0.00-0.00}$
    & $0.97^{+0.01+0.02}_{+0.01-0.16}$
    & $0.71^{-0.04-0.08}_{+0.03+0.19}$
    & $V^{D_s K_{1B}}_0$
    & $0.51^{-0.01+0.03}_{+0.01-0.04}$
    & $0.91^{-0.01+0.00}_{+0.01-0.03}$
    & $0.45^{-0.00-0.07}_{+0.00+0.10}$
    \\
$V^{D_s K_{1B}}_1$
    & $1.50^{+0.01+0.05}_{-0.01-0.07}$
    & $0.59^{-0.02-0.02}_{+0.02+0.03}$
    & $0.10^{-0.01-0.01}_{+0.01+0.02}$
    & $V^{D_s K_{1B}}_2$
    & $-0.12^{+0.01-0.01}_{-0.01+0.01}$
    & $0.68^{-0.02-0.02}_{+0.01+0.01}$
    & $0.36^{+0.01-0.05}_{-0.01+0.08}$
    \\
$A^{D_s  h_{1s}}$
    & $0.10^{+0.00+0.00}_{-0.00-0.00}$
    & $0.93^{-0.00-0.00}_{-0.00-0.00}$
    & $0.51^{-0.03-0.02}_{+0.03+0.02}$
    & $V^{D_s  h_{1s}}_0$
    & $0.57^{-0.01+0.01}_{+0.01-0.01}$
    & $0.89^{-0.01+0.00}_{+0.01-0.00}$
    & $0.37^{+0.00-0.02}_{-0.00+0.03}$
    \\
$V^{D_s  h_{1s}}_1$
    & $1.43^{+0.01+0.02}_{-0.01-0.02}$
    & $0.46^{-0.02-0.01}_{+0.02+0.01}$
    & $0.07^{-0.01-0.00}_{+0.01+0.00}$
    & $V^{D_s  h_{1s}}_2$
    & $-0.17^{+0.01-0.00}_{-0.01+0.00}$
    & $0.55^{-0.02-0.01}_{+0.01+0.01}$
    & $0.20^{-0.01-0.01}_{+0.00+0.01}$
    \\
\hline
$A^{B  b_1}$
    & $0.11^{+0.00+0.00}_{-0.00-0.01}$
    & $1.89^{-0.03-0.03}_{+0.03+0.03}$
    & $1.51^{-0.08-0.09}_{+0.09+0.10}$
    & $V^{B b_1}_0$
    & $0.38^{-0.01+0.03}_{+0.01-0.03}$
    & $1.38^{-0.03+0.00}_{+0.03-0.01}$
    & $0.63^{-0.02-0.04}_{+0.03+0.04}$
    \\
$V^{B b_1}_1$
    & $0.19^{+0.01+0.01}_{-0.01-0.01}$
    & $0.99^{-0.03-0.03}_{+0.03+0.04}$
    & $0.29^{-0.03-0.03}_{+0.03+0.03}$
    & $V^{B b_1}_2$
    & $-0.02^{+0.01-0.01}_{-0.01+0.00}$
    & $1.11^{-20.4+0.33}_{+0.75-0.75}$
    & $7.76^{+59.7-1.82}_{-4.34+3.28}$
    \\
$A^{B  h_{1q}}$
    & $0.10^{+0.00+0.01}_{-0.00-0.01}$
    & $1.89^{-0.03-0.07}_{+0.03+0.07}$
    & $1.51^{-0.08-0.20}_{+0.09+0.25}$
    & $V^{B h_{1q}}_0$
    & $0.37^{-0.01+0.07}_{+0.01-0.06}$
    & $1.37^{-0.03-0.00}_{+0.03-0.03}$
    & $0.62^{-0.02-0.08}_{+0.03+0.11}$
    \\
$V^{B h_{1q}}_1$
    & $0.19^{+0.01+0.02}_{-0.01-0.02}$
    & $0.99^{-0.03-0.08}_{+0.03+0.08}$
    & $0.29^{-0.03-0.06}_{+0.03+0.09}$
    & $V^{B h_{1q}}_2$
    & $-0.02^{+0.01-0.01}_{-0.01+0.01}$
    & $1.11^{-20.4+0.53}_{+0.75-3.74}$
    & $7.76^{+59.7-3.38}_{-4.35+13.7}$
    \\
$A^{B K_{1B}}$
    & $0.12^{+0.00+0.01}_{-0.00-0.01}$
    & $1.78^{-0.03-0.03}_{+0.03+0.04}$
    & $1.27^{-0.07-0.08}_{+0.07+0.11}$
    & $V^{B K_{1B}}_0$
    & $0.45^{-0.01+0.04}_{+0.01-0.04}$
    & $1.37^{-0.03-0.00}_{+0.03-0.01}$
    & $0.54^{-0.02-0.04}_{+0.02+0.05}$
    \\
$V^{B K_{1B}}_1$
    & $0.21^{+0.01+0.01}_{-0.01-0.01}$
    & $0.83^{-0.03-0.03}_{+0.03+0.04}$
    & $0.22^{-0.02-0.02}_{+0.02+0.03}$
    & $V^{B K_{1B}}_2$
    & $-0.05^{+0.01-0.01}_{-0.01+0.01}$
    & $1.74^{-0.08-0.03}_{-0.00+0.02}$
    & $2.17^{-1.09-0.23}_{-0.53+0.31}$
    \\
\hline
$A^{B_s K_{1B}}$
    & $0.08^{+0.00+0.01}_{-0.00-0.01}$
    & $2.06^{-0.04-0.04}_{+0.03+0.05}$
    & $2.57^{-0.20-0.23}_{+0.23+0.33}$
    & $V^{B_s K_{1B}}_0$
    & $0.38^{-0.01+0.04}_{+0.01-0.05}$
    & $1.64^{-0.04-0.03}_{+0.04+0.04}$
    & $1.25^{-0.07-0.13}_{+0.09+0.19}$
    \\
$V^{B_s K_{1B}}_1$
    & $0.15^{+0.01+0.01}_{-0.01-0.02}$
    & $1.34^{-0.05-0.06}_{+0.05+0.07}$
    & $0.76^{-0.08-0.10}_{+0.10+0.14}$
    & $V^{B_s K_{1B}}_2$
    & $-0.06^{+0.01-0.01}_{-0.01+0.01}$
    & $1.65^{-0.01-0.03}_{+0.02+0.04}$
    & $1.16^{-0.02-0.09}_{+0.06+0.13}$
    \\
$A^{B_s  h_{1s}}$
    & $0.09^{+0.00+0.00}_{-0.00-0.00}$
    & $1.95^{-0.04-0.01}_{+0.04+0.01}$
    & $2.11^{-0.16-0.07}_{+0.18+0.07}$
    & $V^{B_s  h_{1s}}_0$
    & $0.51^{-0.01+0.02}_{+0.01-0.02}$
    & $1.60^{-0.03-0.01}_{+0.03+0.01}$
    & $1.05^{-0.06-0.04}_{+0.07+0.04}$
    \\
$V^{B_s  h_{1s}}_1$
    & $0.17^{+0.01+0.00}_{-0.01-0.00}$
    & $1.16^{-0.05-0.02}_{+0.05+0.02}$
    & $0.56^{-0.06-0.03}_{+0.07+0.03}$
    & $V^{B_s  h_{1s}}_2$
    & $-0.10^{+0.01-0.00}_{-0.01+0.00}$
    & $1.52^{-0.02-0.01}_{+0.02+0.01}$
    & $0.95^{-0.03-0.03}_{+0.05+0.03}$
    \\
\end{tabular}
\end{ruledtabular}
\end{table}
}

\subsection{$ B(0^-)\to D^{1/2} ,  D^{3/2}$ Form Factors}

\begin{itemize}

\item From Tables~\ref{tab:B2D}, we notice that most of the
 form factors are small and lie between 0.1 to 0.25, except
$V_0(0)$ and $V_1(0)$ for the transitions emitting $P^{3/2}_1$ states, for
which these lie between 0.5 to 0.6. In contrast with these, $B, B_s \to D, D_s$
form factors carry the highest values between 0.6 to 0.8.

\item Slope parameters carry positive values except $a$ for $V_1$ form factor. However,
these parameters controlling $q^2$ behavior for $V_2$ form factor for
transitions emitting $P^{3/2}_1$ states remains difficult to control in spite of
choosing the $q^2$ dependence given in Eq. (\ref{eq:FFpara}).

\item  Reverse changes occur in the form factors due to the variation in $\beta$
values for initial and final mesons. Increase in $\beta$ for initial (final)
state meson tend to decrease (increase) the magnitude of the form factors,
and

\item Minor changes occur in $B \to D$ form factors
 from their previous values given in the earlier work \cite{Cheng04}, however,
slope parameters show significant difference.

\item To determine the physical form factors for $ B \to D_1$ transitions, one may need
the mixing angle between $D^{1/2}_1$ and $D^{3/2}_1$ states. A mixing angle
$ \theta_{D_1} = (5.76 \pm 2.4)^\circ$ is obtained by Belle through a detailed
$B \to D^* \pi \pi$ analysis \cite{Abe04},  while $ \theta_{D_{s1}} \approx 7^\circ$
is determined from the quark potential model \cite{Cheng03}.

\end {itemize}

\begin{table}[b]
\caption{Form factors of
$ B\rightarrow D^{1/2}_1, D^{3/2}_1$ transitions
obtained in the covariant light-front model are fitted to the
3-parameter form Eq. (\ref{eq:FFpara}). All the form factors are dimensionless.}
\label{tab:B2D}
\begin{ruledtabular}
\begin{tabular}{| c c c c |}
$F$
    & $F(0)$
    & $a$
    & $b$
    \\
\hline
$A^{B  D^{1/2}_1}$
    & $-0.13^{+0.01-0.02}_{-0.01+0.02}$
    & $0.85^{-0.09+0.10}_{+0.08-0.18}$
    & $0.12^{-0.01-0.01}_{+0.02+0.03}$
    \\
$V^{B D^{1/2}_1}_0$
    & $0.11^{-0.01+0.03}_{+0.01-0.03}$
    & $1.08^{-0.02-0.02}_{+0.02-0.07}$
    & $0.08^{-0.03+0.02}_{+0.03-0.04}$
    \\
$V^{B D^{1/2}_1}_1$
    & $-0.19^{+0.02-0.01}_{-0.02+0.01}$
    & $-1.37^{-0.09-0.01}_{+0.08-0.00}$
    & $1.07^{+0.07+0.01}_{-0.06-0.01}$
    \\
$V^{B D^{1/2}_1}_2$
    & $-0.14^{+0.02-0.02}_{-0.02+0.02}$
    & $0.84^{-0.11+0.13}_{+0.09-0.21}$
    & $0.13^{-0.01-0.01}_{+0.02+0.04}$
    \\
\hline
$A^{B  D^{3/2}_1}$
    & $0.25^{-0.01+0.02}_{+0.01-0.02}$
    & $1.17^{-0.03+0.01}_{+0.03-0.03}$
    & $0.33^{-0.02-0.01}_{+0.02+0.01}$
    \\
$V^{B D^{3/2}_1}_0$
    & $0.52^{-0.01+0.04}_{+0.01-0.05}$
    & $1.14^{-0.04+0.02}_{+0.03-0.06}$
    & $0.34^{-0.02-0.01}_{+0.02+0.01}$
    \\
$V^{B D^{3/2}_1}_1$
    & $0.58^{-0.01+0.02}_{+0.01-0.03}$
    & $-0.25^{-0.01-0.03}_{+0.01+0.02}$
    & $0.29^{-0.00+0.01}_{+0.01-0.01}$
    \\
$V^{B D^{3/2}_1}_2$
    & $-0.10^{+0.01-0.02}_{-0.01+0.04}$
    & $-5.95^{-2.07+3.80}_{+1.45-14.67}$
    & $26.2^{+5.8-4.1}_{-11.5+41.0}$
    \\
\hline
$A^{B_s  D^{1/2}_{s1}}$
    & $-0.17^{+0.02-0.02}_{-0.02+0.02}$
    & $0.97^{-0.10+0.06}_{+0.10-0.10}$
    & $0.37^{-0.05-0.04}_{+0.06+0.05}$
    \\
$V^{B_s D^{1/2}_{s1}}_0$
    & $0.13^{-0.01+0.03}_{+0.02-0.03}$
    & $1.14^{-0.04-0.02}_{+0.04-0.05}$
    & $0.29^{-0.04-0.03}_{+0.05+0.04}$
    \\
$V^{B_s D^{1/2}_{s1}}_1$
    & $-0.25^{+0.03-0.01}_{-0.03+0.01}$
    & $-1.20^{-0.10-0.06}_{+0.10+0.07}$
    & $1.02^{+0.07+0.04}_{-0.06-0.05}$
    \\
$V^{B_s D^{1/2}_{s1}}_2$
    & $-0.17^{+0.02-0.02}_{-0.02+0.02}$
    & $0.96^{-0.12+0.08}_{+0.11-0.12}$
    & $0.39^{-0.04-0.04}_{+0.06+0.06}$
    \\
\hline
$A^{B_s  D^{3/2}_{s1}}$
    & $0.24^{-0.01+0.02}_{+0.01-0.02}$
    & $1.26^{-0.06+0.00}_{+0.06-0.02}$
    & $0.60^{-0.06-0.06}_{+0.07+0.06}$
    \\
$V^{B_s D^{3/2}_{s1}}_0$
    & $0.49^{-0.02+0.04}_{+0.02-0.05}$
    & $1.25^{-0.06+0.01}_{+0.05-0.04}$
    & $0.63^{-0.05-0.06}_{+0.06+0.07}$
    \\
$V^{B_s D^{3/2}_{s1}}_1$
    & $0.57^{-0.01+0.03}_{+0.01-0.04}$
    & $-0.11^{-0.02-0.04}_{+0.02+0.04}$
    & $0.32^{-0.01+0.00}_{+0.01+0.01}$
    \\
$V^{B D^{3/2}_{s1}}_2$
    & $-0.09^{+0.01-0.02}_{-0.01+0.03}$
    & $-4.08^{-1.86+2.63}_{+1.24-7.55}$
    & $21.1^{+5.3-8.3}_{-3.6+21.8}$
    \\
\end{tabular}
\end{ruledtabular}
\end{table}

\section{Summary and conclusions}

In this work, we have studied the decay constants and form factors
of the ground-state $s$-wave and low-lying $p$-wave mesons within
a covariant light-front (CLF) approach. In the previous work \cite{Cheng04}, main
ingredients of the CLF quark model were explicitly worked out for both
$s$-wave and $p$-wave mesons. Besides that various form factors of the $D$ and $B$ mesons,
appearing in their transitions to isovector and isospinor $s$-wave and $p$-wave mesons,
were calculated within the framework of the CLF model.
In the present work, we have updated our results for these mesons, and
   extended the analysis to determine the form factors for
       $D_s$ and $B_s$ transitions, and also include the flavor-diagonal isoscalar
       final states. Calculating the decay constants of most of the $s$-wave mesons
        and a few axial vector mesons from the available experimental
        data for various weak or electromagnetic decays, we
        have fixed the shape parameter $\beta$ of the respective mesons, which
        in turn determine the form factors. A few lattice results are
        also used for this purpose. Errors in the $\beta$ parameters are
        fixed from the corresponding experimental errors, otherwise standard
      $10 \%$ uncertainty is assigned to investigate the effects of variation
      in the $\beta$ parameter.
We have then proceeded to obtain the form factors in the CLF quark model for heavy-to-heavy and
heavy-to-light transitions of the charmed and bottom mesons to the
pseudoscalar mesons, vector mesons, scalar mesons and axial vector mesons. The $q^2$ dependence
of the form factors, generally assumed to be
given by Eq. (\ref{eq:FFpara}), is expressed through the slope
parameters, $a$ and $b$. Their sensitivity to the errors and the assigned
uncertainties of the $\beta$ parameters is investigated separately for the initial and the
final mesons.

Our main results are as follows:

\begin{itemize}

\item  For $P \to P$ transitions, $B_s$
 form factors  at $q^2= 0$ are similar to that of the $B$ meson, as if
the spectator quark does not seem to affect them. Particularly, we observe
$F^{B_s D_s} = F^{B D}$, $F^{B_s K}\approx F^{B \pi}$, and
$F^{B_s \eta_s}\approx F^{B \eta_q}$, where
$\eta_{q}= (u\overline{u} + d\overline{d})/\surd {2}$, and
$\eta_{s}$ is pure $(s\overline{s})$ state. To lesser extant, the charmed mesons also show
a similar trend through $F^{D_s K} = F^{D \pi}$ and $F^{D_s \eta_s}\approx F^{D K}$. Heavy-to-light
form factors of the bottom mesons are smaller (around 0.3) than
that of the charmed  mesons, which are around 0.7. The form factor
$F^{PP}_0$ generally shows a monopole behavior, and $F^{PP}_1$
acquires a dipole behavior.

\item For $P \to V$ transitions also, we find
$F^{B_s D^*_s}\approx F^{B D^*}$, $F^{B_s \phi}\approx F^{B \rho}\approx F^{B \omega}$,
 $F^{D_s \phi}\approx F^{D K^*}$ and $F^{D_s K^*}\approx F^{D \rho}\approx F^{D \omega}$,
 where $F$ denotes any of the four form factors, $V, A_0, A_1$ and $A_2$, at $q^2= 0$.
For the bottom mesons, heavy-to-light form factors are
 smaller (from 0.2 to 0.4) than their heavy-to-heavy ones, which lie between 0.6 to 1.
Due to the reliability in fixing the $\beta$ parameters for the $s$-wave
mesons, the form factors at $q^2= 0$ hardly show sensitivity
to the errors in the $\beta$ values, though slope parameters ($a$ and $b$) generally
tend to increase (decrease) with decrease (increase) in $\beta$ for
the initial meson as well as the final meson.

\item Comparing $ P \to P, V$ form factors obtained
here with the results of other works,  BSW model \cite{BSW}, the
Melikhov-Stech (MS) model \cite{Melikhov}, QCD sum rule
(QSR) \cite{Ball91}, light-cone sum rules (LCSR) \cite{LCSR},
lattice calculations \cite{LatticeFF} and perturbative
 QCD approach \cite{RHLi09},  it is found that our form
 factors agree well with the available lattice results, and
 are most close to that of the MS model, except for the $B_s$
 transitions. The LCSR and BSW model results are usually larger
for $P \to V$ form factors for $D$ and $B$ transitions, whereas
the QSR and pQCD calculations are generally lower than our results.

\item For $P \to S$ transitions, we have calculated the form factors
involving heavy scalar mesons only. These form factors, though are smaller
than the corresponding $P \to P$ form factors, also satisfy
$F^{B_s K^*_0} = F^{B a_0} = F^{B f_{0q}}$,
$F^{B_s f_{0s}} \approx F^{B D^*_0} = F^{B K^*_0}$, and
$F^{D_s f_{0s}}\approx F^{D a_0} = F^{D f_{0q}}$.
All the bottom meson form factors, lying between 0.25 and 0.30,
 are roughly half of that of the charmed mesons, which are
 around 0.5-0.6. The suppression of the $D_0^{*0}\pi^-$
production relative to $D^0\pi^-$ one clearly favors a smaller
$B\to D_0^*$ form factor relative to the $B\to D$ one.
The form factor $F^{PS}_0$ shows a monopole behavior, and $F^{PS}_1$
has a dipole behavior in general. The $P \to S$ form factors
are found to increase slowly with $q^2$ compared to the $P \to P $
ones. For the bottom sector,
the slope parameters are larger in magnitude than that for the charm sector.
These parameters (except for the case of negative $a$)
show an increase (decrease) with decrease (increase) in $\beta$ for each
of the initial and final mesons.

\item For heavy-to-light  $P \to A(1^{++})$ transitions, the form
factor $A^{PA}(0)$ for the bottom mesons, generally lying around 0.25,
is larger than that for charmed meson transitions for which it
lies close to 0.16. Similarly, the form factor $V^{PA}_2(0)$ is $<0.1$
for the charm sector, where as it lies around 0.2 for the
 bottom transitions. In contrast, the form factor $V^{PA}_1(0)$,
  lying around 0.4 for the bottom sector is significantly
  smaller than that  for the charm sector, where its value
   lies between 1.4 and 1.8. Also $V^{PA}_0(0)$ for the bottom
   transitions is roughly half of its value for the charmed
    meson transitions. We also observe that
 $V^{PA}_1 > V^{PA}_0 > A^{PA} > V^{PA}_2$ for the charmed mesons
 and $V^{PA}_1 > A^{PA} > V^{PA}_2 > V^{PA}_0$ for the bottom mesons.
 All the slope parameters, except
    for $V^{PA}_1$ and $V^{PA}_2$ for the charmed meson
 transitions, are found to be positive, and for the
  bottom mesons, their values are significantly
  larger than that for the charm sector.

\item For heavy-to-light transitions $P \to A(1^{+-}) $,
the form factors, $A^{PA}(0), V^{PA}_0(0)$ and $V^{PA}_1(0)$, are positive,
 where as $V^{PA}_2(0)$ is negative and small (around -0.1) for
  the bottom as well as the charmed sector. These follow the pattern:
 $V^{PA}_1 > V^{PA}_0 > A^{PA} > |V^{PA}_2| $ for the charmed mesons
 and  $V^{PA}_0 > V^{PA}_1 > A^{PA} \geq |V^{PA}_2| $  for the bottom mesons.
  Numerically, the form factor
$A^{PA}(0)$ is generally around 0.1, where as $V^{PA}_0(0)$ lies close
 to 0.5 for all the cases. Form factor $V^{PA}_1(0)$, lying between
  0.15 to 0.20, for the bottom sector is significantly smaller than
   that for the charm sector, where its value lies between
   1.3 to 1.6. Typically for the heavy-to-light $P \to A(1^{+-})$ transitions,
   the form factors $A^{PA}(0)$ and $V^{PA}_1(0)$ usually increase (decrease) with
 increasing (decreasing) beta for initial meson as well as for
  final meson. Both $V^{PA}_0(0)$ and $V^{PA}_2(0)$ form factors
   decrease (increase) in magnitude
with increasing (decreasing) $\beta$ for the initial mesons, and show the opposite
trend for the final mesons. For the bottom sector,
the slope parameters are found to be larger than that for the charm sector.

\item For $B \to D^{1/2}_1/D^{3/2}_1$ transitions, all the
 form factors lie between 0.1 to 0.2, except for
$V_0(0)$ and $V_1(0)$ for the transitions emitting $P^{3/2}_1$
states, for which these lie between 0.5 to 0.6. Slope parameters
carry positive values, except
 $a$ for the $V_1$ form factor. Reverse changes occur in the
 form factors due to variation in $\beta$ values for
  initial and final mesons, i.e., increase in $\beta$ for the initial (final)
state meson tend to decrease (increase) the magnitude of the form factors.

\item Now several model calculations for
  $B\rightarrow A$ form factors are available: the ISGW2 model \cite{ISGW2},
  the constituent quark-meson model (CQM) \cite{Deandrea},
 the QCD sum rules (QSR) \cite{Aliev}, light cone sum rules
 (LCSR) \cite{KCYang08}, and the
 perturbative QCD (pQCD) approach \cite{RHLi09}.
 Significant differences are observed, since these approaches differ in
  their treatment of dynamics of the form factors.
  For instance, $V^{Ba_1}_0 = 1.20$, obtained in the CQM model, and $1.01$ in the ISGW2 model,
   is much larger than its values obtained in other approaches. The BaBar
   and Belle measurements \cite{BABAR, Abe07} of
 $\overline{B}^0\rightarrow a^{\pm}_1 \pi^{\mp}$  seem to favor a
 value of $V^{Ba_1}_0\approx 0.30$ \cite{ChengAP}. We have earlier pointed out
\cite{Cheng04} that relativistic effects could manifest
in heavy-to-light transitions at maximum recoil where the final-state meson can be
highly relativistic, which can naturally be considered in the CLF model.
Various form factors, calculated using the CLF model,
have earlier been used to study weak hadronic and radiative decays
of the bottom mesons emitting $p$-wave mesons \cite{Cheng07, Cheng10},
 and a good agreement between theory and available experimental
data could be obtained. It has been pointed out in the previous work \cite{Cheng04} that
the requirement of HQS is also satisfied for
the decay constants and the form factors obtained in the CLF quark model. Particularly, it has been
shown that the Bjorken \cite{Bjorken} and Uraltsev \cite{Uraltsev} sum rules for the
Isgur-Wise functions are satisfied.

\end{itemize}

\vskip 2.5cm \acknowledgments  RCV thanks
the Institute of Physics, Academia Sinica and National Center for
Theoretical Sciences, National Tsing-Hua University, for their hospitality
during his visits, where most part of this work was done. He specially acknowledges
H.Y. Cheng for the invitation, useful discussions and reading the manuscript.
He also expresses his thanks to C.K. Chua for providing his source codes. This research 
was supported in part by the National Science Council of R.O.C. under 
No. NSC97-2112-M-001-004-MY3 and by the special NCTS grant of Academia Sinica.

\eject

\end{document}